\documentclass[preprintnumbers,article,amsmath,amssymb,floatfix,10pt,prd,twocolumn,
superscriptaddress,nofootinbib]{revtex4-2}
\usepackage{bm}
\usepackage{amsfonts}
\usepackage{latexsym}
\usepackage{graphicx}
\usepackage{amsmath}
\usepackage{palatino}
\usepackage{mathpazo}
\usepackage{textcomp}
\usepackage{booktabs}
\usepackage{dcolumn}
\usepackage{ragged2e}
\usepackage{hyperref}
\hypersetup{colorlinks,citecolor=blue}
\hypersetup{colorlinks=true,linkcolor=red,filecolor=magenta,    urlcolor=blue}
\usepackage{amsmath}
\usepackage{xcolor}
\usepackage{orcidlink}
\usepackage{epsfig}
\usepackage{caption}
\usepackage{xcolor}
\usepackage{subcaption}
\usepackage{commath}
\usepackage{etoolbox}
\makeatletter
\patchcmd{\@citex}{\if@filesw\immediate\write}{\if@filesw\immediate\write\@gobble}{}{}
\makeatother

\def\jnl@style{\it}
\def\aaref@jnl#1{{\jnl@style#1}}

\def\aaref@jnl#1{{\jnl@style#1}}

\def\aj{\aaref@jnl{AJ}}                   % Astronomical Journal
\def\apj{\aaref@jnl{ApJ}}                 % Astrophysical Journal
\def\apjl{\aaref@jnl{ApJ}}                % Astrophysical Journal, Letters
\def\apjs{\aaref@jnl{ApJS}}               % Astrophysical Journal, Supplement
\def\apss{\aaref@jnl{Ap\&SS}}             % Astrophysics and Space Science
\def\aap{\aaref@jnl{A\&A}}                % Astronomy and Astrophysics
\def\aapr{\aaref@jnl{A\&A~Rev.}}          % Astronomy and Astrophysics Reviews
\def\aaps{\aaref@jnl{A\&AS}}              % Astronomy and Astrophysics, Supplement
\def\mnras{\aaref@jnl{Mon.~Not.~Roy.~Astron.~Soc.}}             % Monthly Notices of the RAS
\def\prd{\aaref@jnl{Phys.~Rev.~D}}        % Physical Review D
\def\prc{\aaref@jnl{Phys.~Rev.~C}}  % Physical Review C
\def\prl{\aaref@jnl{Phys.~Rev.~Lett.}}    % Physical Review Letters
\def\qjras{\aaref@jnl{QJRAS}}             % Quarterly Journal of the RAS
\def\skytel{\aaref@jnl{S\&T}}             % Sky and Telescope
\def\ssr{\aaref@jnl{Space~Sci.~Rev.}}     % Space Science Reviews
\def\zap{\aaref@jnl{ZAp}}                 % Zeitschrift fuer Astrophysik
\def\nat{\aaref@jnl{Nature}}              % Nature
\def\aplett{\aaref@jnl{Astrophys.~Lett.}} % Astrophysics Letters
\def\apspr{\aaref@jnl{Astrophys.~Space~Phys.~Res.}} % Astrophysics Space Physics Research
\def\physrep{\aaref@jnl{Phys.~Rep.}}      % Physics Reports
\def\physscr{\aaref@jnl{Phys.~Scr}}       % Physica Scripta
\def\commat{\aaref@jnl{Comm.~Math.~Phys.}}              % Communications in Mathematical Physics
\def\science{\aaref@jnl{Science}}               % Science
\def\cqg{\aaref@jnl{Classical Quant.~Grav.}}            % Classical and Quantum Gravity
\def\jpcs{\aaref@jnl{JPCS}}                                     % Journal of Physics Conference Series
\def\ijmpd{\aaref@jnl{Int.~J.~Mod.~Phys.~D}}                    % International Journal of Modern Physics D
\def\grg{\aaref@jnl{Gen.~Relat.~Gravit.}}               % General Relativity and Gravitation
\def\rpp{\aaref@jnl{Rep.~Prog.~Phys.}}          % Reports on Progress in Physics
\def\npa{\aaref@jnl{Nucl.~Phys.~A}}        % Nuclear Physics A
\def\lrr{\aaref@jnl{Living Rev.~Rel.}}                   % Living reviews in relativity
\def\jcap{\aaref@jnl{J.~Cosmology Astropart.~Phys.}}    % Journal of cosmology and astroparticle physics
\def\rmp{\aaref@jnl{Rev.~Mod.~Phys.}}   %Reviews of modern physics
\def\epjc{\aaref@jnl{Eur.~Phys.~J.~C}} 
\def\plb{\aaref@jnl{~Phy.~Lett.~B}} 
\def\mpla{\aaref@jnl{Mod.~Phy.~Lett.~A}} 
\def\arxiv{\aaref@jnl{arxiv.org}}

\allowdisplaybreaks[1]
\renewcommand{\arraystretch}{1.1}
\begin{document}
%\color{red}
\color{black}       %% For one column
%
%\title{Cosmological Bouncing Solutions in $f(Q)$ Gravity}
\title{Generalized Ghost Pilgrim Dark Energy Fractal Cosmology with Observational Constraint}
%\end{document}

\author{S.R. Bhoyar 
\orcidlink{0000-0001-8427-4540}}
\email{drsrb2014@gmail.com}
\affiliation{Department of Mathematics, Phulsing Naik Mahavidyalaya Pusad-445216 Dist. Yavatmal (India)}

\author{Yash B. Ingole\orcidlink{0009-0006-7208-1999}}
\email[Corresponding author: ]{ingoleyash01@gmail.com}
\affiliation{Department of Mathematics, Phulsing Naik Mahavidyalaya, Pusad-445216, Dist. Yavatmal, India}

\author{A.P. Kale 
\orcidlink{0009-0005-0253-7965}}
\email{akshaykale1000@gmail.com}
\affiliation{Department of Mathematics, B. Raghunath ACS College, Parbhani-431401 (India)}
%
%%%%%%%%%%%%%%%%%%%%%%%%%%%%%%%%%%%%%  DATE  %%%%%%%%%%%%%%%%%%%%%%%%%%%%%%%%%%%%

\begin{abstract}

In this work, we explore dark energy models, mainly ghost, generalized ghost, and generalized ghost pilgrim dark energy models within the framework of fractal cosmology. To obtain solutions for the field equations, we employed a parameterization of the deceleration parameter, as proposed by \textit{R. K. Tiwari}. By utilizing  Markov Chain Monte Carlo (MCMC) analysis, we impose constraints on the free parameters of the derived solutions. The analysis is based on observational datasets, including 57 data points from the Observational Hubble Data ($OHD$) and, 1048 points from the $Pantheon$ Supernovae sample. This approach allows us to assess the viability of the dark energy models in describing the current cosmic expansion. According to the effective equation-of-state parameter, the model maintains itself in the quintessence era and ultimately switches into the Einstein-de Sitter model. Furthermore, we investigate the statefinder, jerk, snap, and lerk parameters. The energy conditions of each model satisfy the weak and null energy conditions but violate the strong energy condition. We find that the $Om(z)$ curves for the data samples exhibit a consistently negative slope throughout the entire range.\\


\end{abstract}

\maketitle

\date{\today}

%%%%%%%%%%%%%%%%%%%%%%%%%%%%%%%%%%%%%%%%%%%%%%%%%%%%%%%%%%%%%%%%%%%%%%%%
%%%%%%%%%%%%%%%        Introduction        %%%%%%%%%%%%%%%%%%%%%%%%%%%%%
%%%%%%%%%%%%%%%%%%%%%%%%%%%%%%%%%%%%%%%%%%%%%%%%%%%%%%%%%%%%%%%%%%%%%%%%
\section{Introduction} 
Extensive research has been conducted to broaden our knowledge of dark-driven, late-time cosmic expansion. Our universe is expanding at an accelerated rate, which is revealed by evidence gathered from the high-z Supernovae team via direct observations \cite{r1,r2}. Solheim first observed this state of accelerated universe expansion \cite{r3}. A non-vanishing cosmological constant and a negative deceleration parameter are features of the model that best fit the facts, according to observations of the luminosities of many galaxy clusters. Additionally, the claim has received much attention due to recent advancements in type-Ia Supernovae \cite{r4,r5,r6}, CMBR anisotropy \cite{r7,r8}, large-scale structure \cite{r9,r10}, baryon acoustic oscillation (BAO) \cite{r11}, and weak lensing \cite{r12}. These discoveries have led to the observation that our current universe is geometrically flat moreover, the most unexpected and revolutionary conclusion from these findings is that out of the total energy of the universe, 4.6\% is consumed by nonrelativistic or baryonic matter, and 24\% of it is a nonbaryonic matter known as dark matter (DM). The remaining 71.4\% is observed as an unknown form of exotic cosmic fluid with negative pressure, called dark energy (DE). DE generates a strong negative force that leads to an antigravity effect, which causes acceleration, despite gravitational attraction \cite{r13,r14,r15}. Numerous theoretical and observational studies have been conducted to investigate the composition of DE. The primary issue in this case is that, because DE does not interact with baryonic matter, it is impossible to find this bizarre element. DE also refers to an isotropic fluid in general relativity (GR) with a constant energy density and negative pressure. On the basis of various observational findings, cosmologists suggest that the cosmological constant \( \Lambda \) is the leading candidate for dark energy (DE). This is because, throughout cosmic evolution, it maintains a constant energy density with negative pressure. However, despite being considered, this idea of dark energy has faced challenges, particularly the fine-tuning and cosmic coincidence problems \cite{r16}. Consequently, different models of dynamically evolving DE with an effective equation-of-state (EoS) parameter \( \omega \) have been proposed in numerous studies. Nevertheless, the precise value of the EoS of DE remains unknown. Our lack of insight enables us to recommend many DE candidates. In addition to the $\Lambda$CDM model, which takes into account the cosmological constant with $\omega=-1$, other canonical scalar field models that have been suggested as potential solutions for the DE problem include quintessence $(- 2/3 \leq \omega \leq - 1/3)$ \cite{r17,r18}, phantom field $(\omega < -1)$ \cite{r19}, k-essence \cite{r20,r21}, tychons \cite{r22}, quintom \cite{r23}, and the interacting DE candidates Chaplygin gas \cite{r24,r25}, holographic models \cite{r26,r27,r28}, and brane world \cite{r29,r30} have been offered as varieties of feasible solutions for the DE problem. 

As previously mentioned, the most popular and straightforward model for understanding dark energy within the framework of the $\Lambda$CDM model is the cosmological constant $(\Lambda)$. However, effective in describing the universe's accelerating expansion, the $(\Lambda)$ concept has many important theoretical and observational limitations, such as the fine-tuning problem \cite{r31}, the coincidence problem \cite{r32,r33}, lack of dynamical explanation \cite{r34}, challenges in quantum gravity \cite{r35}. These limitations motivate us to search for alternative dark energy models, such as ghost dark energy (GDE) and pilgrim dark energy (PDE). GDE emerged from attempts to resolve the dark energy mystery through the lens of quantum field theory (QFT) in curved spacetime. Specifically, it is connected to the concept of the Veneziano ghost \cite{r36,r37,r38,r39,r40}, which arises to solve the U(1) problem in the low-energy effective theory of quantum chromodynamics (QCD) \cite{r41,r42,r43,r44,r45,r46}. It is introduced in gauge theories to cancel unphysical degrees of freedom. In Minkowski spacetime, the ghost field does not affect the vacuum energy density; however, in curved spacetime, it generates a small vacuum energy density such that $\rho_{DE} \propto H\Lambda^3_{QCD}$, where $H$ is the Hubble parameter and $\Lambda_{QCD}$ is the mass scale of order $100MeV$. This provides the correct magnitude $\sim (3\times 10^{-3}eV)^4$ for the observed dark energy density with $H\sim 10^{-3}eV$. This extraordinary numerical coincidence also implies that the fine-tuning problem is eliminated in this model \cite{r36,r37,r38,r39,r40}. This new model has an advantage over other dark energy models in that it does not require the introduction of any additional parameters, degrees of freedom, or modifications to gravity because it is fully integrated into the standard model and general relativity. Numerous studies have been conducted on the dynamic behavior of the ghost dark energy (GDE) model in both flat \cite{r47} and nonflat \cite{r48} universes. The energy density of the GDE is considered proportional to the Hubble parameter in all the aforementioned investigations. However, $H + O(H^2)$ is often the energy density of the Veneziano ghost field in QCD \cite{r49}. While the subleading term $H^2$ may have had a significant role in the early evolution of the universe, only the leading component $H$ has been considered thus far for the energy density of GDE \cite{r50}. Compared with the standard GDE model, the inclusion of the second term in the energy density of GDE results in better agreement with observations \cite{r51}. We refer to this model as generalized ghost dark energy (GGDE), as in \cite{r52}. The formula for the energy density of the GGDE model is $\rho_{DE}=\xi H +\eta H^2$, where $\xi$ and $\eta$ are the model constants. The GGDE model has demonstrated the ability to produce a de Sitter phase of expansion as well as the phantom regime of expansion $(\omega_d<-1)$ in the presence of dark matter and dark energy interaction \cite{r52}. The other properties of the GGDE model are described in \cite{r53,r54}.

On the other hand, the holographic dark energy (HDE) model is a popular and fascinating model in quantum gravity and GR. It is based on Cohen et al. \cite{r55} relationship between ultraviolet and infrared cutoffs, which states that a system's size ($L$) should be less than the mass of a black hole (BH). To prevent BH formation, the bound of energy density could be violated. To prevent BH formation, a strong repulsive force is needed. Recent observations have shown that the repulsive force provided by $(\omega<-1/3)$ and $(\omega>-1)$ is insufficient to prevent BH formation. Therefore, phantom-like DE $(\omega<-1)$ is the most suitable repulsive force to prevent BH formation. However, the mass of BH is decreased by phantom-like deformation (DE) \cite{r56,r57}. The 50\% reduced area phantom scalar field accreted into a BH \cite{r58}. In phantom DE, the mass of the BH becomes zero before the enormous rip singularity \cite{r59}. In the early universe, the phantom energy lowers the mass of the BH \cite{r60}. Despite all the discussions and observations, this issue remains unresolved. Such circumstances motivate Wei \cite{r61} to construct a new dark energy model the pilgrim dark energy (PDE) model which is an extension of the generalized Chaplygin gas model. He hypothesized that the phantom DE possesses a strong force that repels the formation of BH. Additionally, he placed constraints on this model by using the Hubble horizon as an example. Before this idea, there was a great deal of discussion on the evolution of HDE models that incorporate various IR cutoffs both in the FRW and KK universes \cite{r62,r63,r64,r65,r66,r67,r68,r69}, both with and without interaction with cold dark matter (CDM). Furthermore, many cosmological parameters have been examined to enhance the dependability of these models. To differentiate these models from the $\Lambda$CDM model, particularly $\omega-\omega'$ \cite{r70,r71,r72,r73} and state-finders analysis \cite{r74,r75,r76,r77,r78} have been studied. Now the GGDE has been reformulated via the PDE as $\rho=(\xi H +\eta H^2)^\psi$, where $\psi$ is constant; this modified gravity is known as generalized ghost pilgrim dark energy (GGPDE) \cite{r79,r80}. This concept arises from the need to address the limitations of these models (GGDE and PDE) by combining the benefits of both and introducing further generalizations. GGPDE seeks to provide a more flexible and comprehensive framework that can better explain the universe's accelerated expansion. These ghost fields contribute to the vacuum energy density, which is hypothesized to play the role of dark energy in the universe. Research on this topic is thoroughly discussed and cited in references \cite{r81,r82,r83,r84}, providing detailed insights and relevant findings that contribute to the understanding of the subject.

To address the complexity of dark energy models and the need for strict assumptions about underlying physics, cosmographic methods have become valuable tools in recent DE studies. These methods allow researchers to conduct model-independent analyses of cosmic expansion using observational data alone. By focusing on cosmographic parameters like jerk, snap, and lerk, it is possible to explore the universe's evolution without relying on specific gravitational frameworks, such as fractal cosmology or general relativity. This approach informs our study, where we apply cosmography to investigate fractal DE models, aiming to provide an observationally grounded evaluation of DE dynamics.

We extend our study to solve the fractal space-time problem in the FRW universe via the deceleration parameter $q=\lambda-(\mu/H)$ suggested by \cite{r85,r86}. We explored three dark energy models (GDE, GGDE, and GGPDE), and we derived an effective equation of states by maintaining the relationship of the energy densities of these dark energy models with the Hubble parameter. We employ the $R^2$ minimization strategy to fine-tune the model and find the best-fit values for the parameters $\lambda$, and $\mu$. By comparing the theoretical predictions with observational data, we may determine which parameter set best fits the empirical evidence with the help of statistical analysis. The dataset includes observational Hubble data ($OHD$) data samples and $Pantheon$ samples. By offering a wide range of observable constraints, these datasets enable us to thoroughly test and confirm our cosmological models.

The paper is structured as follows: Section \eqref{section 2} consists of a concise overview of fractal cosmology, including its mathematical formulation. In sect. \eqref{section 3}, we derive the deterministic solution of the field equation with varying deceleration parameters. To restrict the model parameters, in Sect.\eqref{section 4}  we employ 57 Hubble data points and 1048 $Pantheon$ data points. Error bar charts were used to compare the model to the standard $\Lambda$CDM model.
Section \eqref{section 5} is dedicated to the study of all cosmographic parameters such as deceleration, effective EoS, Statefinder, jerk, snap, and lerk parameters and $Om(z)$ diagnosis. In Sect \eqref{section 6}, we determine the energy conditions of dark energy models to help assess their physical viability and compatibility with the fundamental laws of general relativity. Finally, we discuss the results and conclusions of our study in Sect. \eqref{section 7}.

%In these models, a parametrized form of DE density is considered and relying upon the requirements, the parametrized form is converted to produce viable models.The likelihood of $\omega \ll -1$ has been confirmed by recent cosmological evidence from Supernovaee 1a \cite{asti/2006}, CMB \cite{mact/2006}, and large-scale structures \cite{koma/2009}. On the other hand, the DE that crosses the phantom dividing line ($\omega \ll -1$) is marginally preferred. A few other limits derived from observational data [10, 34] are, respectively, $-1.67 < \omega < -0.62$ and $-1.33 < \omega < -0.79$. As of right now, all we know about DE is that it is spatially homogenous and non-clustered; while its influence was negligible in the beginning, it currently rules the universe.

\section{Overview and Mathematical Formulation of the DE Models in the Fractal cosmology}
\label{section 2}
Fractal cosmology is a framework that challenges the traditional assumption of large-scale homogeneity in the universe. Unlike the standard cosmological model, which is based on the cosmological principle that posits that the universe is uniform and isotropic on large scales, fractal cosmology suggests that the universe exhibits self-similar, scale-invariant structures across different scales. This concept draws inspiration from fractal geometry, a mathematical framework introduced by Benoit Mandelbrot \cite{r87}, which describes objects that exhibit patterns that repeat at multiple magnification levels, i.e., the universe seems the same from every galaxy, on the basis of a conditional cosmological principle in a fractal universe. In fractal cosmology, the distribution of galaxies, clusters, and large-scale structures is not homogeneous but instead follows a fractal pattern \cite{r88}. Observations of galaxy clustering and voids support the idea that these structures may not simply smooth out at larger scales, but continue to exhibit complexity and self-similarity. This approach challenges conventional theories of cosmic structure formation, which rely on the gravitational collapse of matter in a nearly uniform universe \cite{r89,r90}. The fractal model of the universe has gained attention because of its ability to describe the large-scale inhomogeneities observed in galaxy distributions, without invoking exotic components such as dark matter or dark energy. By focusing on scale-invariance and the fractal dimension of cosmic structures, this approach offers an alternative explanation for the observed distribution of matter in the universe, potentially addressing some of the limitations of the $\Lambda$CDM model. However, fractal cosmology remains controversial. Critics argue that the universe appears to become homogeneous on enormous scales, as indicated by large-scale surveys such as the sloan digital sky survey (SDSS) \cite{r91}. Neverthless this, fractal cosmology continues to be explored as a possible avenue for understanding cosmic structure and evolution, especially in light of unresolved issues in standard cosmology.

Linde \cite{r92} proposed a chaotic inflationary universe model that exists indefinitely and explains fractal cosmology. To study power-counting renormalizable field theory in a fractal space, Asghari et al. \cite{r93} investigate the fractal model of the world. RA El-Nabulsi et al., \cite{r94} have researched the quantum cosmology paradigm using fractal mathematics and fractional Dunkl Laplacian. The fractional-fractal Friedmann and Raychaudhuri equations for an isotropic and homogeneous universe are examined by P. F. da Silva Junior et al., \cite{r95}. They also explored how Padmanabhan's idea fits within the framework of fractional quantum gravity. Through applying the first law of thermodynamics and the derivation of the Friedmann equations, R. Jalalzadeh et al. \cite{r96} shed light on how such a fractal universe evolves. The cosmological framework suggested by EW. Costa et al. \cite{r97} seeks to answer the problem of synchronicity with respect to the universe's age. It is based on fractional quantum cosmology. They have created a novel fractional $\Lambda$CDM cosmological model and acquired the required formalism by deriving the fractional Hamiltonian constraint in a general minisuperspace to accomplish this. D. Pawar et al. examined perfect fluid anisotropic plane symmetric space-time within the context of fractal cosmology \cite{r98}. A flat FRW cosmological model with wet dark fluid has also been researched on the basis of fractal gravity \cite{r99}. A. Siyal and K.A. Abro proposed a newly developed free convection Newtonian fluid model \cite{r100}, and numerically examined via both fractal and fractional methods, which are based on the fractal mathematical model of the Caputo differential operator. The fractal FRW cosmological model, which includes two types of dark energy, was examined by D. Pawar et al. \cite{r101}. They used the DE that is allowed in our solar system to study the behavior of the universe in a fractal framework.

The total action of Einstein's gravity in fractal space-time is given by \cite{r102,r103}
\begin{equation}\label{eqI}
    S=S_G +S_M,
\end{equation}
where $S_G$ and $S_M$ are the gravitational and matter parts of the action, respectively, which are given by
\begin{equation}\label{2}
    S_G =\frac{1}{16\pi G} \int d\tau(x) \sqrt{-g}(R-2\Lambda -\phi \partial_uv \partial^uv),  
\end{equation}

\begin{equation}\label{3}
    S_M = \int d\tau(x) \sqrt{-g} L_M,
\end{equation}
where $G$, $\Lambda$, and $R$ represent the gravitational constant, cosmological constant, and Ricci scalar respectively, and where $g$ is the determinant of the dimensionless metric $g_{uv}$. Furthermore, the fractal parameter is $\phi$, and the fractional function is $v$. A Lebesgue-Stieltjes measure $d\tau(x)$ is substituted for the normal measure $d^4x$ of action, and $L_M$ is a matter Lagrangian The Friedmann equations in a fractal world are as follows: Taking the variation of action [Eq.\eqref{eqI}] about the FRW metric $g_{uv}$, we obtain

\begin{equation}\label{4}
    H^2 + H \frac{\Dot{v}}{v} + \frac{k}{a^2} -  \frac{1}{6}\phi \Dot{v}^2 = \frac{8 \pi G}{3} \rho_{DE} + \frac{\Lambda}{3},
\end{equation}
\begin{equation}\label{5}
\begin{split}
    \Dot{H} +H^2 -H \frac{\Dot{v}}{v} + \frac{1}{6}\phi \Dot{v}^2 -\frac{1}{2}\frac{\square v}{v}=-\frac{8\pi G}{3} ( \rho_{DE}  &+  3P_{DE})\\
    &\ + \frac{\Lambda}{2},
    \end{split}
\end{equation} 
where $\square=\partial_{u}\partial^u$, $H=\Dot{a}/a$ is the Hubble parameter, and where $\rho_{DE}$, and $P_{DE}$ are the energy density and pressure of the DE model respectively.
The curvature parameters $k=0,1$, and $-1$ correspond to the flat, closed, and open universes respectively. The overdot $(.)$ represents the derivative concerning the cosmic time $t$. In the present work, we consider $k=0$ as the flat universe. The continuity equation in a fractal universe takes the form

\begin{equation}\label{6}
    \Dot{\rho}_{DE}+\bigl(3H+\frac{\Dot{v}}{v}\bigr)(\rho_{DE}+P_{DE})=0.
\end{equation}

The conventional Friedmann equations are recovered for $v = 1$. Furthermore, we assume that spatial slices have a regular geometry and that only the time direction is fractal. Indeed, classically fractals can be spacelike $v=v(x)$ or even timelike $v=v(t)$ within the paradigm of fractal cosmology (see Ref. \cite{r103} for details). There is no quantum equivalent of space-like or time-like fractals because these two scenarios lead to different classical physics, but at the quantum level all configurations should be taken into account. We consider a timelike fractal in this work. As a result, the parts that are related to $x$ remain unchanged, whereas the parameters that depend on time change.

We assume a time-like fractal profile $v =t^{-\beta}$, where $\beta=4(1-\alpha)$ (with $0< \alpha \leq 1$ ) is the fractal dimension. The Friedmann equations given in \eqref{4} and \eqref{5} in the  absence of the cosmological constant can be written as

\begin{equation}\label{7}
       H^2 - H \frac{\beta}{t} + \frac{k}{a^2} - \frac{1}{6}\frac{\phi \beta^2}{t^{2(\beta+1)}} = \frac{8 \pi G}{3} \rho_{DE},
\end{equation}

\begin{equation}\label{8}
\begin{split}
       \Dot{H} +H^2 -\frac{1}{2}\frac{\beta}{t}H + \frac{\beta(\beta+1)}{2t^2} & +\frac{\phi\beta^2}{3t^{2(\beta+1)}}=\\
       &\  -\frac{8\pi G}{3} (\rho_{DE} +3P_{DE}).
       \end{split}
\end{equation}

While the continuity equation given in \eqref{6} takes the form

\begin{equation}\label{9}
    \Dot{\rho}_{DE}+\bigl(3H-\frac{\beta}{t}\bigr)(\rho_{DE}+P_{DE})=0
\end{equation}

The fractional integral definition and fractal calculus stated in \cite{r103,hilfer2000applications} explain that $\alpha$ ranges between 0 and 1. For $\alpha = 1$, we obtained $\beta = 0$, which indicates no fractal structure in the universe. This allows us to recover the Friedmann equations used in standard cosmology. From Friedmann equations \eqref{7} and \eqref{8}, it is clear that a time-like fractal profile lacks a limit as \( t \to 0 \) because the equations diverge unless \( \beta = 0 \), implying that the early stages of the universe cannot have a timelike fractal structure.

\section{Solution of the field equations with varying deceleration parameter}
\label{section 3}
The Hubble parameter $(H)$ and the deceleration parameter $(q)$ are crucial for interpreting how the cosmos has evolved under different cosmological scenarios. The Hubble parameter provides information on the current rate of expansion, whereas the deceleration parameter reflects whether the universe is accelerating $(q < 0)$ or decelerating $(q > 0)$. Here, we define these parameters.
\begin{equation}\label{10}
    H=\frac{\Dot{a}(t)}{a(t)},
\end{equation}

\begin{equation}\label{11}
    q=- \frac{\Dot{H}}{H^2}-1,
\end{equation} where $a(t)$ is the scale factor. R.K. Tiwari et al. \cite{r85,r86},  assumed that $H$ and $q$ were inversely related. The following is an expression for this relationship, assuming that $\lambda$ and $\mu$ are constants:
\begin{equation}\label{12}
    q=\lambda -\frac{\mu}{H}.
\end{equation}
provides a simplified yet effective way to model the evolution of  \( q \) in terms of  \( H \) within the framework of FLRW cosmology. A variable deceleration parameter is considered because it aligns with the universe's transition from a decelerating phase in the past to the accelerating phase we observe today. This transition is supported by recent astrophysical evidence, including observations of Type Ia Supernovae \cite{riess1998observational,perlmutter1999measurements} and cosmic microwave background (CMB) anisotropies\cite{spergel1997age}. This form is particularly useful in fractal gravity because it captures essential dynamical features of cosmic evolution without resorting to a complex equation of state. These derivations can reveal the dynamical effects of fractal gravity and further clarify whether the model can replicate observed cosmic acceleration or even predict novel behavior unique to fractal settings.

By comparing Eqs. \eqref{11} and \eqref{12}, we can derive the following expression for $H$

\begin{equation}\label{13}
    H= \frac{\mu}{c \mu e^{-\mu t} +\lambda+1},
\end{equation}
where $c$ is an integrating constant. Following the integration of the above equation, we obtain the scale factor $a(t)$ as

\begin{equation}\label{14}
    a=k_1 \big[c \mu +(\lambda+1) e^{\mu t}, \big]^{\frac{1}{\lambda+1}}
\end{equation}
where $k_1$ denotes an integrating constant. Notably, selecting $c=-\frac{(\lambda+1)}{\mu}$  leads to a point type singularity at $t = 0$. This decision allows for the estimation of the Hubble parameter $(H)$ and scale factor $a(t)$

 \begin{equation}\label{15}
     H= \frac{\mu e^{\mu t}}{(e^{\mu t}-1)(\lambda+1)},
 \end{equation}

 \begin{equation}\label{16}
     a(t) = (e^{\mu t}-1)^{\frac{1}{\lambda+1}}.
 \end{equation}

 The deceleration parameter $(q)$ [Eq. \eqref{11}] in terms of cosmic time $(t)$ [Eq. \eqref{15}] is expressed as

 \begin{equation}\label{17}
     q=-1+ \frac{(1+\lambda)}{e^{\mu t}}.
 \end{equation}

In our model, when $t=\frac{1}{\mu}\ln{[\lambda+1]}$, the deceleration parameter's sign changes. Furthermore, we may determine the relationship between the cosmic time $(t)$ and redshift $(z)$ as follows by connecting the redshift $z$ and the universe's scale factor  $a(t)=(1+z)^{-1}$.

\begin{equation}\label{18}
    t=\frac{1}{\mu}\ln\big[{(1+z)^{-(1+\lambda)}+1}\big].
\end{equation}

Now, the Hubble parameter $(H)$ is expressed in terms of redshift $(z)$ as

\begin{equation}\label{19}
    H(z)=\frac{\mu}{1+\lambda}[(1+z)^{1+\lambda}+1].
\end{equation}

The behavior of these cosmological parameters is now being investigated statistically. The next stage combines the observed Hubble data values of $H(z)$ to obtain the best-fit values for the model parameters $\Lambda$ and $\mu$.

 \section{Observational data}
\label{section 4}
We employ various datasets to constrain the parameters of our cosmological model. These datasets include $Pantheon$ Supernovae samples, baryon acoustic oscillation (BAO) datasets, and cosmic chronometer (CC) datasets. The $Pantheon$ Supernovae samples provide a total of $1048$ data points, whereas the BAO and CC datasets contribute 6 and 57 data points, respectively. To analyze the data, we utilize the emcee python library, which implements the Markov chain Monte Carlo (MCMC) method for Bayesian analysis and estimation of the likelihood function. This approach allows us to explore the parameter space effectively and obtain reliable constraints on our cosmological model.

\subsubsection{\textbf{Observational Hubble Datasets ($OHD$)}}

The Hubble parameter is also related to the differential
redshift as $H(z) =-\frac{1}{1+z} \frac{dz}{dt} $, where $dz$ is obtained from the spectroscopic surveys, so a measurement of $dt$ provides
the Hubble parameter, which is independent of the model. Two methods are generally used to measure
the Hubble parameter values $H(z)$ at certain redshift and are extraction of $H(z)$ from line-of-sight BAO data and differential age (DA) estimation of H(z). We notice the observational constraints on the parameters $\mu$ and $\lambda$ using the latest 57 data points of $H(z)$ in the redshift range, $0.07\leq z\leq 2.4$, in which 31 points are obtained via DA method, whereas  26 points are obtained via BAO. The expression contains two model parameters, $\mu$ and  $\lambda$. To find the cosmological parameters, we need to find some approximate values of those model parameters. We consider the Observational Hubble Datasets ($OHD$) to obtain some best-fit values of these model parameters. We find the best fit curve of $H(z)$ with 57 observed values, via the following statistical formula $R^2-test$:
\begin{equation}\label{20}
	R^2=1-\frac{\sum_{1}^{57}[(H_i)_{obs}-(H_i)_{th}]^2}{\sum_{1}^{57}[(H_i)_{obs}-(H_i)_{mean}]^2}
\end{equation}
where $(H_i)_{obs}-(H_i)_{th}$ is the difference between the observed value and the theoretical value obtained from the best-fit plot.

\subsubsection{\textbf{Apparent Magnitude, Luminosity Distance}}\label{E12}
The expansion of the universe has been supported by the $SNeIa$ observation. The $SNeIa$ data are recorded from the Panoramic Survey
Telescope and Rapid Response System (Pan-STARSS1), the
Sloan Digital Sky Survey (SDSS), the Supernovae Legacy
Survey (SNLS), and the Hubble Space Telescope (HST) survey \cite{scolnic2018complete}.

The total flux of the source of light is measured by the luminosity distance, which is defined as:\\
\begin{center} 
	$d_L(z)=(1+z)\int_{0}^{z}\frac{H_0}{H(z^*)}dz^*.$
\end{center}
To obtain the best-fit values of the model parameters $n$ and $H_0$, we need to define the apparent magnitude in terms of $d_L$.
\begin{center} 
	$m(z) =M+5\log_{10}d_L-5\log_{10}\big(\frac{H_0}{Mpc}\big)+25,$
\end{center}
where $M$ is constant for all $SNeIa$. 
The distance modulus $\mu(z)=m-M$ is given by
%\begin{center} 
	\begin{equation}\label{21} 
	\mu(z) =5\log_{10}d_L-5\log_{10}\big(\frac{H_0}{Mpc}\big)+25.
\end{equation}
%\end{center}
To obtain the best-fit curve of apparent magnitude $m(z)$ with the use of the $pantheon$ sample dataset of
1048 points of distance moduli $\mu_z$
towards the range
$0.01 \leq z_i \leq 2.26$ for various redshifts \cite{scolnic2018complete}. From \cite{suzuki2012hubble} the statistically significant value of $M$ is -19.30.

To perform the best curve fitting of the theoretical and observed results, we used the $R^2-test$ formula for both datasets:
\begin{equation*}
    {\large R^2=1-\frac{\sum_{1}^{1048}[(\mu_i)_{obs}-(\mu_i)_{th}]^2}{\sum_{1}^{1048}[(\mu_i)_{obs}-(\mu_i)_{mean}]^2}\\}.
\end{equation*}

Now, from the equations of $d_L(z)$, $m(z)$ and  $\mu(z)$, for $z=-1$, they are indefinite. We require $ -1< z$ and $\lambda \neq-1$, therefore, we find the best fit values of $\mu$ and $\lambda$ by restricting the parametric space $ -1<z$ and $\lambda \neq-1$.

The error bar plots of the 57 points of the $OHD$  and the 1048 points of  $Pantheon$ datasets are plotted in Figs. \eqref{1a} and \eqref{2a} via the constrained values of the model parameters as shown in Tab. \ref{tab I}. Both datasets are compared  with the well-accepted $\Lambda$ CDM model (dotted black line) for $H_{0} = 67.8$$ km/s/Mpc$, $\Omega_{\Lambda_{0}} = 0.7$ and $\Omega_{m_{0}} = 0.3$ as shown in Figs. \eqref{1a} and \eqref{2b}. The maximum likelihood contours for the model parameters $n$ and $H_0$ are shown in the following Figs. \eqref{1b} and \eqref{2b} for independent Hubble and   $pantheon$ datasets with $1-\sigma$ and $2-\sigma$ error contours in the $H_0-n$ plane.

\begin{widetext}

\begin{figure}[!h]
\centering
\begin{subfigure}[b]{0.45\textwidth}
\centering
\includegraphics[width=\textwidth]{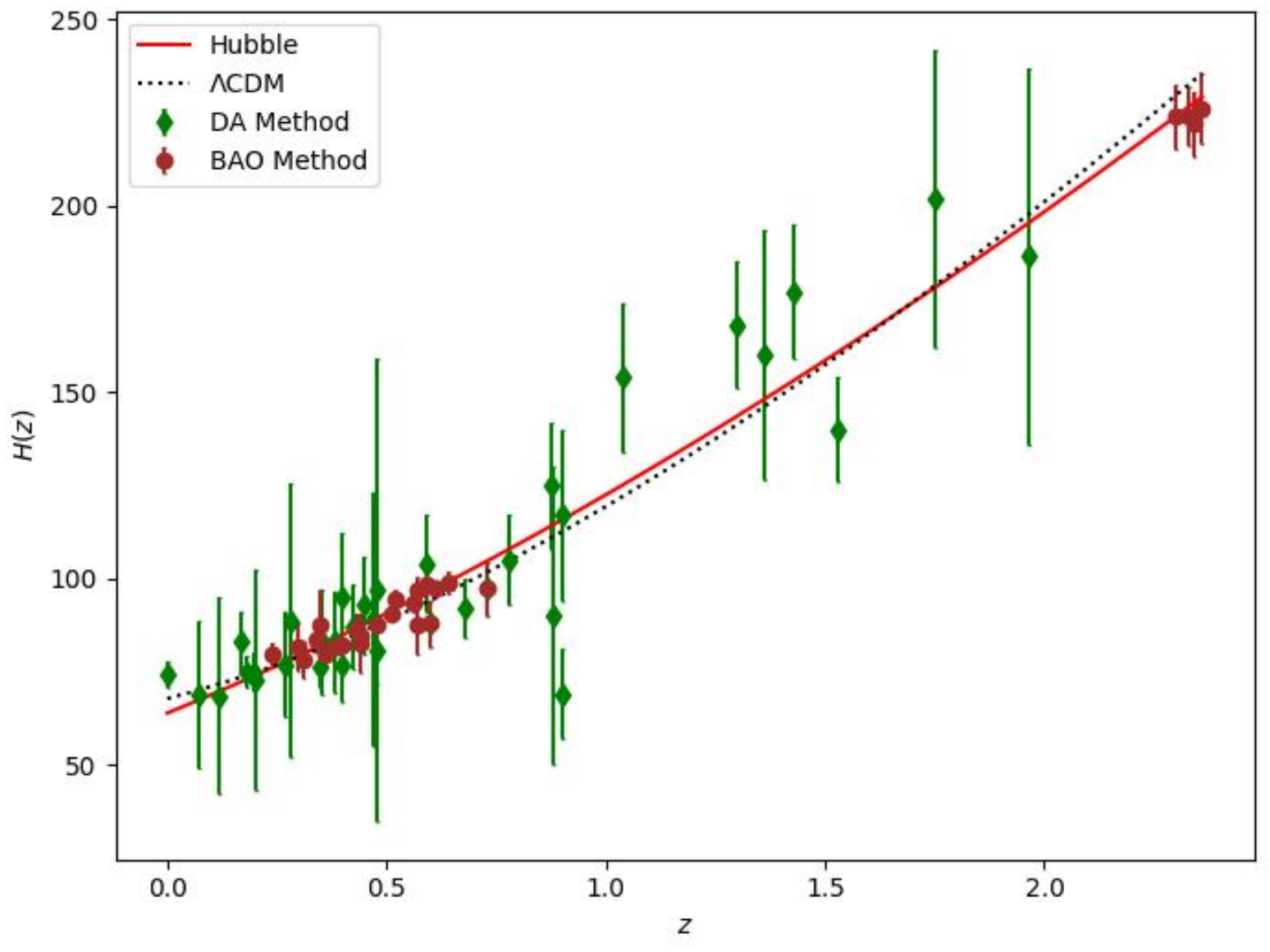}\caption{\justifying{The error bar plots for 57 data points from Hubble datasets together with best-fit plots.}} \label{1a}
\end{subfigure}
\hfill
\begin{subfigure}[b]{0.47\textwidth}
        \centering
        \includegraphics[width=\textwidth]{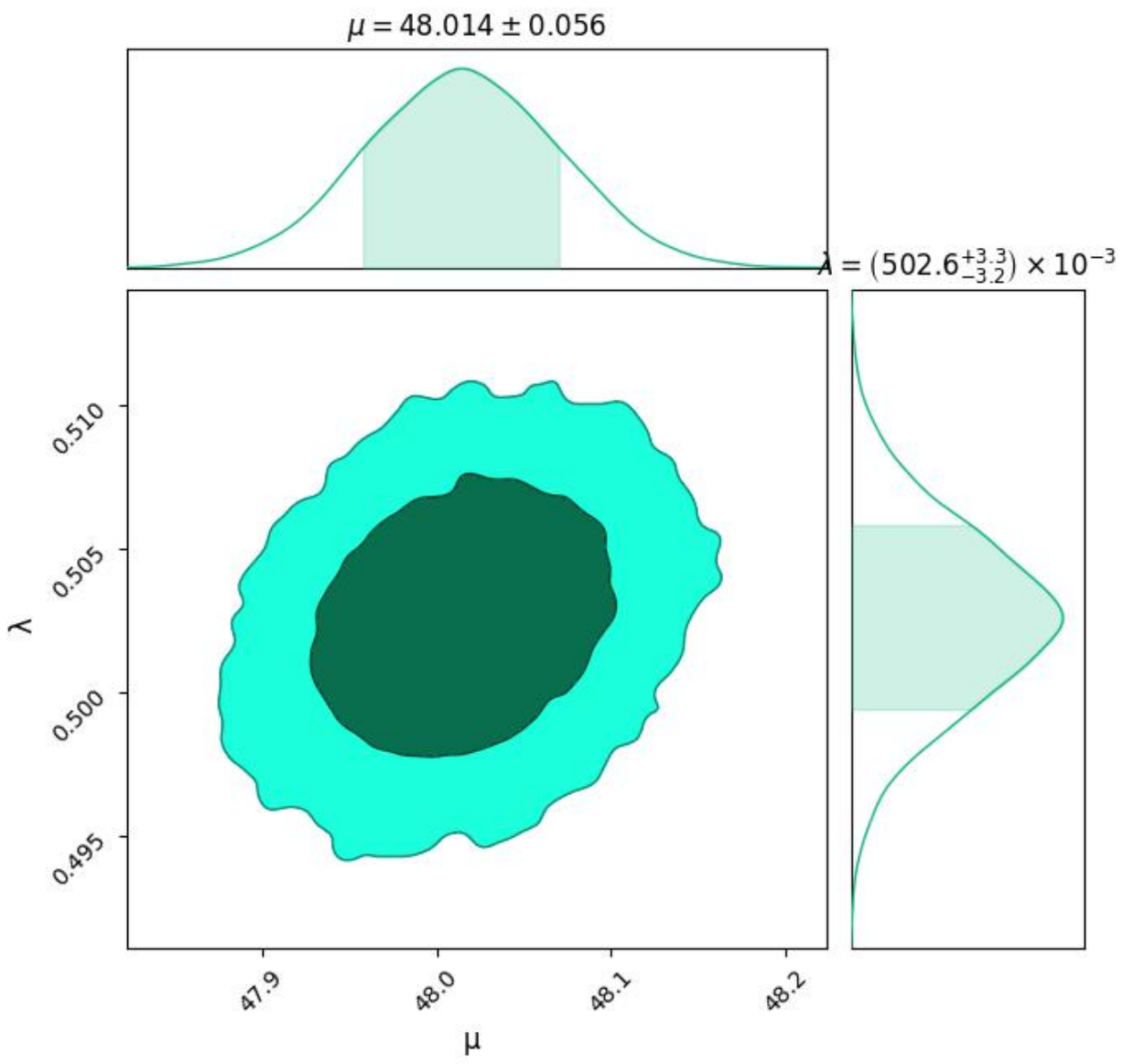}
\caption{Contour plots for Hubble dataset $H(z)$.} \label{1b}
    \end{subfigure}
\caption{}
    \label{Fig. 1}
    \end{figure}
%\end{widetext}

%\begin{widetext}
\begin{figure}[!h]
\centering
\begin{subfigure}[b]{0.45\textwidth}
\centering
\includegraphics[width=\textwidth]{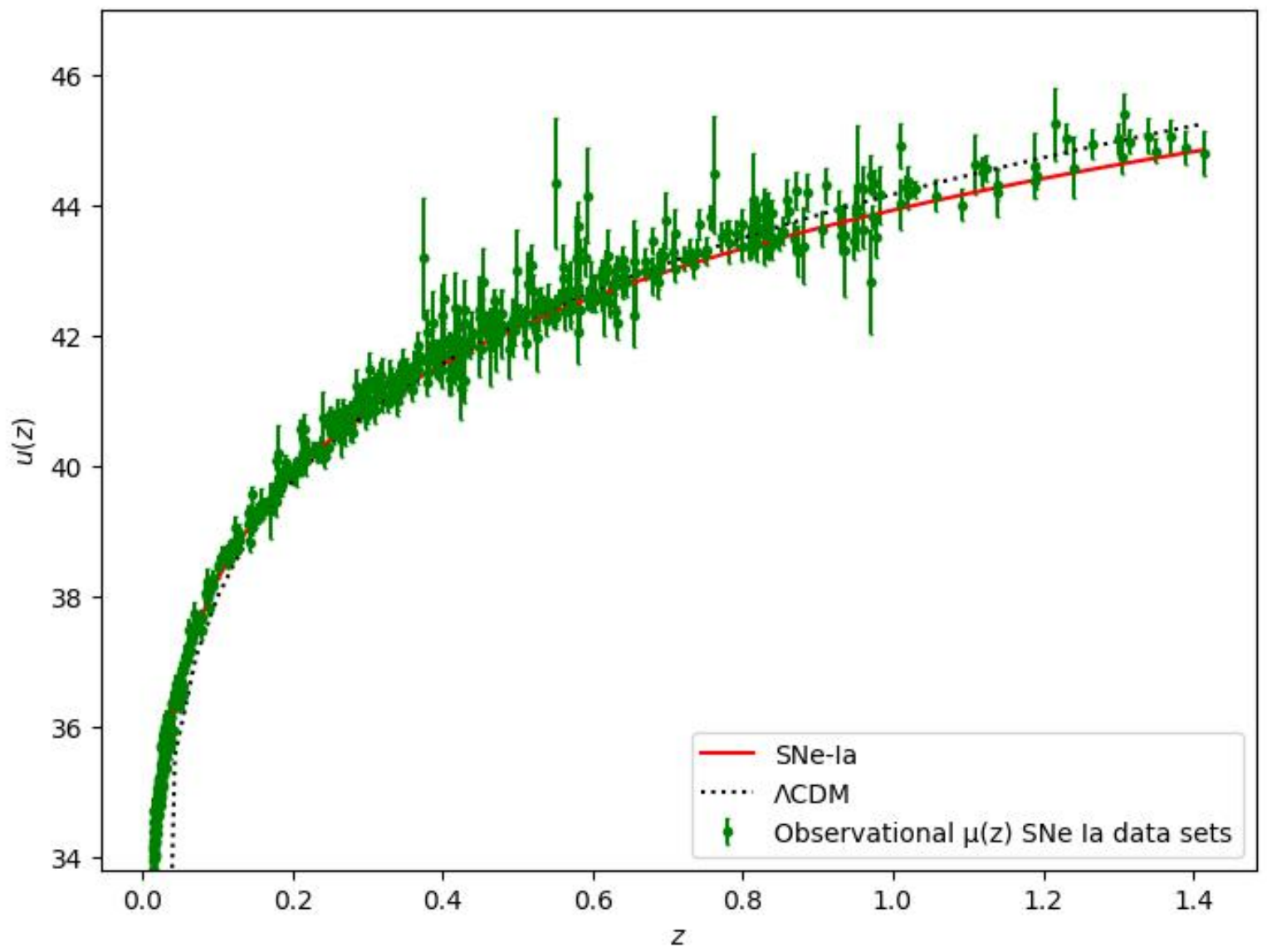}\caption{\justifying{The error bar plots for 1048 data points from $Pantheon$ datasets together with best-fit plots.}} \label{2a}
\end{subfigure}
\hfill
\begin{subfigure}[b]{0.47\textwidth}
        \centering
        \includegraphics[width=\textwidth]{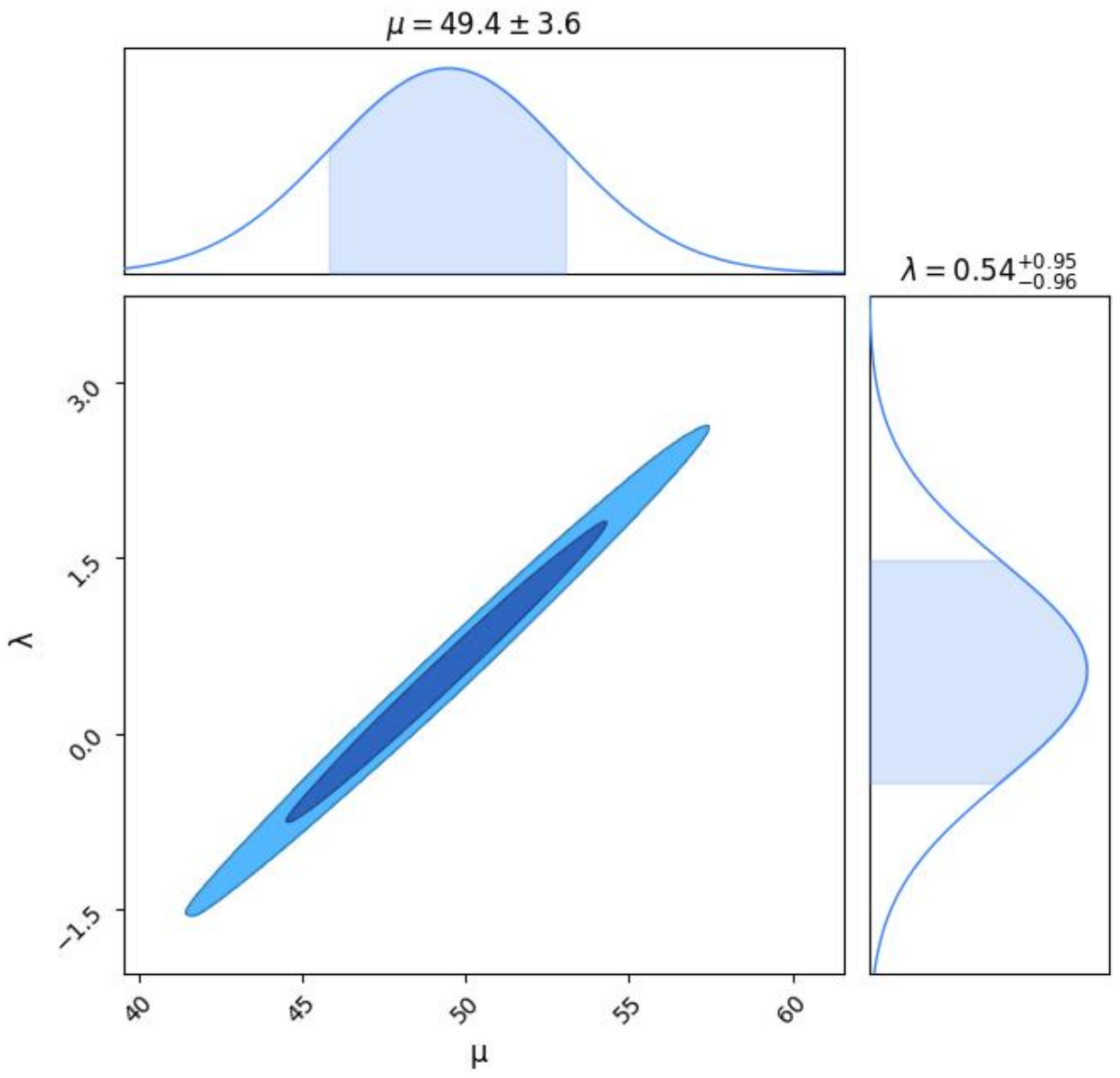}
\caption{Contour plots for $Pantheon$ dataset.} \label{2b}
    \end{subfigure}
\caption{}
    \label{Fig. 2}
    \end{figure}
   \end{widetext}
%\end{figure}
%\end{widetext}

Since the energy density of the GDE is directly proportional to the Hubble parameter, i.e., $\rho_{GDE}=\xi H$ \cite{r40}, where $\xi$ is constant along with the equation of continuity \eqref{9}, conversion of cosmic time $(t)$ into redshift $(z)$ \eqref{18}, and Hubble parameter value \eqref{19}, the energy density and pressure for the GDE model are given as

\begin{widetext}
    
\begin{equation}\label{22}
    \rho_{GDE}(z)= \frac{\mu \xi}{1+\lambda}\big[(1+z)^{1+\lambda}+1 \big],
\end{equation}

\begin{equation}\label{23}
%\begin{split}
 P_{GDE}(z)  =  \frac{\mu \xi \big[(1+z)^{1+\lambda} +1 \big]}{1+\lambda} \bigg\{\frac{\ln\big[{(1+z)^{-(1+\lambda)}+1}\big]}{\ln\big[{(1+z)^{-(1+\lambda)}}+1\big]^3} 
         \frac{\big[(\lambda -2 )(1+z)^{1+\lambda}-3\big]+4(1-\gamma)(1+\lambda)}{ \big[(1+z)^{1+\lambda} +1 \big]-4(1-\gamma)(1+\lambda)}\bigg\},
        %\end{split}
\end{equation}

\end{widetext}

where $\Dot{H}=\frac{d}{dt}H=-(1+z)H(z)\frac{d}{dz}H(z)$. However, to understand the evolution of the early universe, Cai et al. \cite{r47} introduced the generalized ghost DE density, which includes the subleading term $H^2$ in the ordinary ghost DE, i.e., $\rho_{GGDE}=\xi H + \eta H^2$, where $\eta$ is a constant with dimension $[energy]^2$. When Eqs. \eqref{9}, \eqref{18}, and \eqref{19} are employed, the energy density and pressure for the GGDE model are represented as
\begin{widetext}
\begin{equation}\label{24}
%\begin{split}
    \rho_{GGDE}(z)=\frac{\mu\big[(1+z)^{1+\lambda}+1\big]}{(1+\lambda)^2}\big\{ \xi (1+\lambda)+ \eta\mu[(1+z)^{1+\lambda}+1]\big\},
    %\end{split}
\end{equation}

%\begin{widetext}
\begin{equation}\label{25}
\begin{split}
     P_{GGDE}(z) = & \frac{(1+z)^{1+\lambda}+1}{3(1+\lambda)^2\big\{(1+z)^{1+\lambda} \ln[{(1+z)^{-(1+\lambda)} +1}]^\frac{1}{\mu}-4(1-\gamma)(1+\lambda)\big\}} \Bigg\{(1+\lambda)(1+z)^{1+\lambda} \ln\big[\\
     &\ {(1+z)^{-(1+\lambda)}+1}\big] \big[\xi (1+\lambda)+2\eta\mu [(1+z)^{1+\lambda}+1]\big] - \big[\xi (1+\lambda)+\eta\mu((1+z)^{1+\lambda}+1)\big]\\
     &\ \big[[(1+z)^{1+\lambda}+1] \ln[{(1+z)^{-(1+\lambda)}+1}]^3 -4(1-\gamma)(1+\lambda)\big]\Bigg\}.
\end{split}
\end{equation}
\end{widetext}

Moreover, the GGPDE model is an extension of GDE in the PDE model whose energy density is defined as $\rho_{GGPDE}=(\xi H + \eta H^2)^\psi$, where $\psi$ is the PDE parameter \cite{r79,r80}. We reform the equation of energy density and pressure for the GGPDE model with Eqs. \eqref{9}, \eqref{18}, and \eqref{19} as follows:

\begin{widetext}

\begin{equation}\label{26}
%\begin{split}
\rho_{GGPDE}(z) =\Big[\ \frac{\mu[(1+z)^{1+\lambda}+1]}{(1+ \lambda)^2}\Big] ^\psi  \bigg\{\xi(1+\lambda)
 +\eta\mu\big[(1+z)^{1+\lambda}+1\big] \bigg\},
%\end{split}
\end{equation}

\begin{equation}\label{27}
P_{GGPDE}(z)=\Big[\frac{\mu[(1+z)^{1+\lambda}+1]}{(1+\lambda)^2}\Big] ^\psi \Bigg\{ \frac{\big\{\xi(1+\lambda)+\eta\mu\big[(1+z)^{1+\lambda}+1\big]\big\}^\psi \big\{\xi(1+\lambda)+2\eta\mu\big[(1+z)^{1+\lambda}+1\big]\big\}}{
\begin{split}
\big\{\big[(1+z)^{1+\lambda}+1\big] \ln\big[{(1+z)^{-(1+\lambda)}  +1\big]^3}- & 4(1-\gamma)(1+\lambda)\big\}\big\{\xi(1 +\lambda)+\\
&\ \eta\mu\big[(1+z)^{1+\lambda}+1\big]\big\}\Bigg\}.
\end{split}
}
\end{equation}
\end{widetext}

\section{Cosmographic Parameters}
\label{section 5}

\subsection{Deceleration parameter}

With the help of Eqs. \eqref{17} and \eqref{18}, the deceleration parameter $(q)$ is given as 

\begin{equation}\label{28}
    q(z) = \frac{\lambda(1+z)^{1+\lambda}-1}{(1+z)^{1+\lambda}+1}.
\end{equation}

\begin{figure}[!h]
\centering
\includegraphics[scale=0.7]{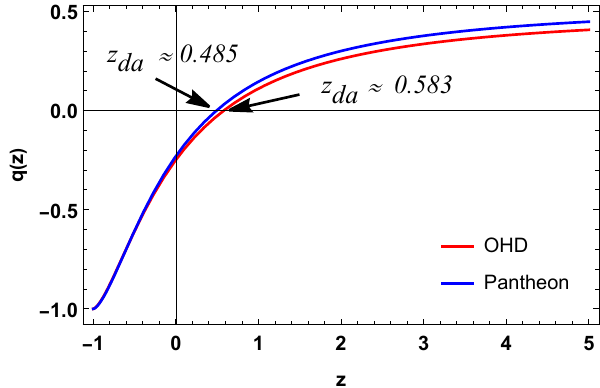}
\caption{Plot of deceleration parameter $(q)$ versus redshift $(z)$, for the $OHD$ and $pantheon$ datasets.}
\label{Fig.5}
\end{figure}

Understanding the expansion dynamics of the universe and its fundamental elements requires examining the behavior of the deceleration parameter $(q)$. We can distinguish between different cosmological models and gain insights into the dominance of matter, dark energy, and dark matter at various stages of cosmic evolution. Therefore, it is essential to
investigate the behavior of the deceleration parameter ($q$).

The positive value of $(q)$ represents slowing i.e., decelerating and the negative value represents speeding i.e., accelerating expansion of the universe. Fig. \eqref{Fig.5} shows the behavior of the deceleration parameter $(q)$ associated with the model parameter $\lambda$ constrained by the $OHD$ and $Pantheon$ data. The model transitions from early deceleration to late acceleration. At the present epoch, the given dataset shows the values of the deceleration parameter as $q_0=-0.249^{+0.002}_{-0.001}$ and $q_0=-0.230^{+0.475}_ {-0.480}$; from these values, it is clear that initially, the model struggles to attempt the widely accepted value,i.e., $q_0 \approx- 0.55$, but later the model efficiently enters an accelerating de Sitter regime, characterized by $q=-1$ \cite{r104}. The deceleration-acceleration transition values of the redshift $z$ are $z_{da}=0.583\pm 0.437$ and $z_{da}=0.485 \pm 0.111$, which are closer to the widely accepted standard value $z_{da} \approx 0.7$ shown by recent observations \cite{r105}. 

\subsection{EoS parameters}
The effective equation of state $(\omega_{eff})$ is the ratio of pressure to density with this equation we derived the equation of effective EoS for the GDE, GGDE, and GGPDE models.
\begin{widetext}
\begin{equation}\label{29}
    \omega_{GDE}(z)= -1 + \frac{\ln\big[{(1+z)^{-(1+\lambda)} +1}\big]^{\mu}\big[(1+z)^{1+\lambda} +1 \big](1+ \lambda)(1+z)^{1+\lambda}}{\mu\big[(1+z)^{1+\lambda}+1\big]\bigg\{\ln\big[{(1+z)^{-(1+\lambda)} +1}\big]^3 \big[(1+z)^{1+\lambda}+1\big]-4(1-\gamma)(1+\lambda)\bigg\}},
\end{equation}

\begin{equation}\label{30}
\begin{split}
       \omega_{GGDE}(z)= -1+ \frac{\ln\big[{(1+z)^{-(1+\lambda)}+1}\big]^\mu\big\{\xi (1+\lambda)+2\eta\mu\big[(1+z)^{1+\lambda}+1\big]\big\}(1+z)^{1+\lambda}}{\big\{\xi(1+\lambda)+\eta\mu^2\big[(1+z)^{1+\lambda} +1\big]\big\}\big\{ \big[(1+z)^{1+\lambda }+1\big]\ln\big[{(1+z)^{-(1+\lambda)}+1}\big]^3-4(1-\gamma)(1+\lambda)\big\}}, 
\end{split}
\end{equation}

\begin{equation}\label{31}
    \omega_{GGPDE}(z)=-1 + \frac{\psi(1+z)^{1+\lambda}\ln\big\{{(1+z)^{-(1+\lambda)}+1}\big\}\big\{\xi(1+\lambda)+2\eta\mu\big[(1+z)^{1+\lambda}+1\big]\big\}}
    {\begin{split} 
    \big\{\big[(1+z)^{1+\lambda}+1\big]\ln\big[{(1+z)^{-(1+\lambda)}+1\big]^3}-4  (1-\gamma) (1+\lambda)\big\}& \big\{\xi(1+\lambda)+\\
    &\ 
 \eta\mu\big[(1+z)^{1+\lambda}+1\big]\big\}.
    \end{split}}
\end{equation}

\end{widetext}

%\begin{widetext}
      
\begin{figure}[h!]
\centering
\begin{subfigure}[b]{0.31\textwidth}
\centering
\includegraphics[width=\textwidth]{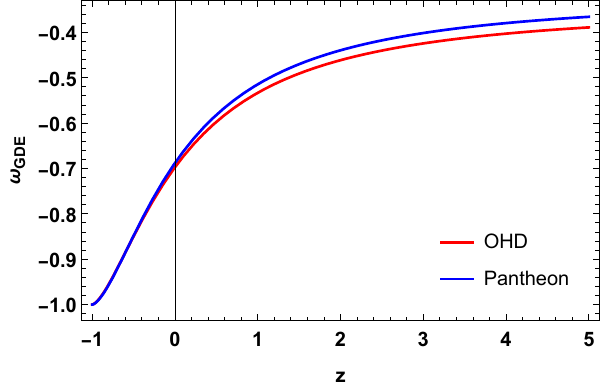}\caption{GDE model.} \label{Fig. 6a}
\end{subfigure}
\hfill
\begin{subfigure}[b]{0.31\textwidth}
        \centering
        \includegraphics[width=\textwidth]{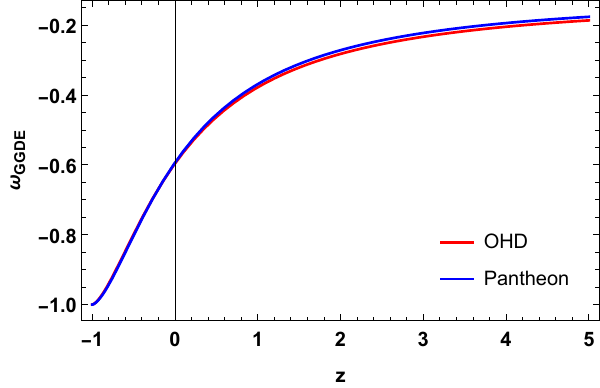}
\caption{GGDE model.} \label{Fig. 6b}
    \end{subfigure}
    \hfill
\begin{subfigure}[b]{0.31\textwidth}    \centering
\includegraphics[width=\textwidth]{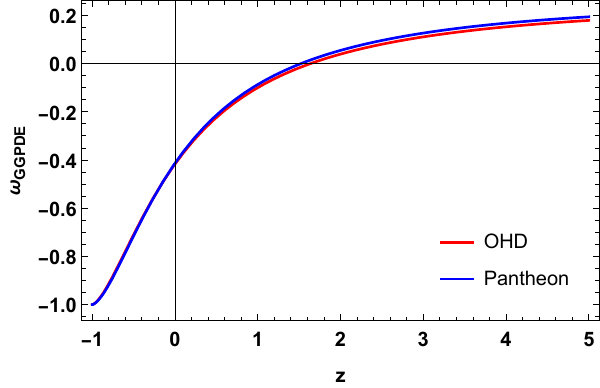}
\caption{GGPDE model.} \label{Fig. 6c}
\end{subfigure}
\caption{Plot of the effective EoS parameter $(\omega_{eff})$  versus the redshift $(z)$ for the $OHD$ and $pantheon$ datasets with constrained values of the model parameters $\lambda$ and $\mu$ of the GDE, GGDE, and GGPDE models, respectively. }
    \label{Fig. 6}
\end{figure}

%\end{widetext}

The universe's expansion has been classified by the effective EoS parameter. When $\omega_{eff}=1$, it represents a stiff fluid. $\omega_{eff}=0$ denotes the matter-dominated phase, whereas $\omega_{eff}=1/3$ indicates the radiation-dominated phase. Conversely, the quintessence phase is indicated by the relation $-1< \omega_{eff} \leq 0$ and the cosmological constant, or $\Lambda$CDM model, is shown by $\omega_{eff}=-1$. When $\omega_{eff}<-1$, the phantom era is observed. In an accelerating universe, $\omega_{eff}$ should be less than $-1/3$.

The behavior of the effective EoS parameter for the GDE, GGDE, and GGPDE models, constrained by observational data, is presented in Fig. \eqref{Fig. 6}. Notably, at present, the values of $\omega_{eff}$ fluctuate between $-0.4$ and $-0.6$ for each dataset, indicating that the models lie within the quintessence regime. Table \eqref{tab II} provides detailed values of $\omega_0$ for each model. Finally, the models are expected to behave like a cosmological constant in the near future, i.e., $\omega=-1$.

%\begin{widetext}

%\end{widetext}

\subsection{Statefinder parameters}

This method is typically employed to examine different dark energy (DE) models and gain insight into their characteristics via higher-order derivatives of the scale factor. The statefinder diagnostic pair $\{r, s\}$, being dimensionless, allows for the analysis of DE cosmic properties without depending on specific models and can be calculated through established expressions \cite{r74,r75,r76,r77,r78}.

\begin{equation}\label{32}
    r=2q^2+q+(1+z) \frac{dq}{dz},
\end{equation}

\begin{equation}\label{33}
    s=\frac{r-1}{3(q-\frac{1}{2})}.
\end{equation}

\begin{figure}[!h]
\centering
\begin{subfigure}[b]{0.45\textwidth}
\centering
\includegraphics[width=\textwidth]{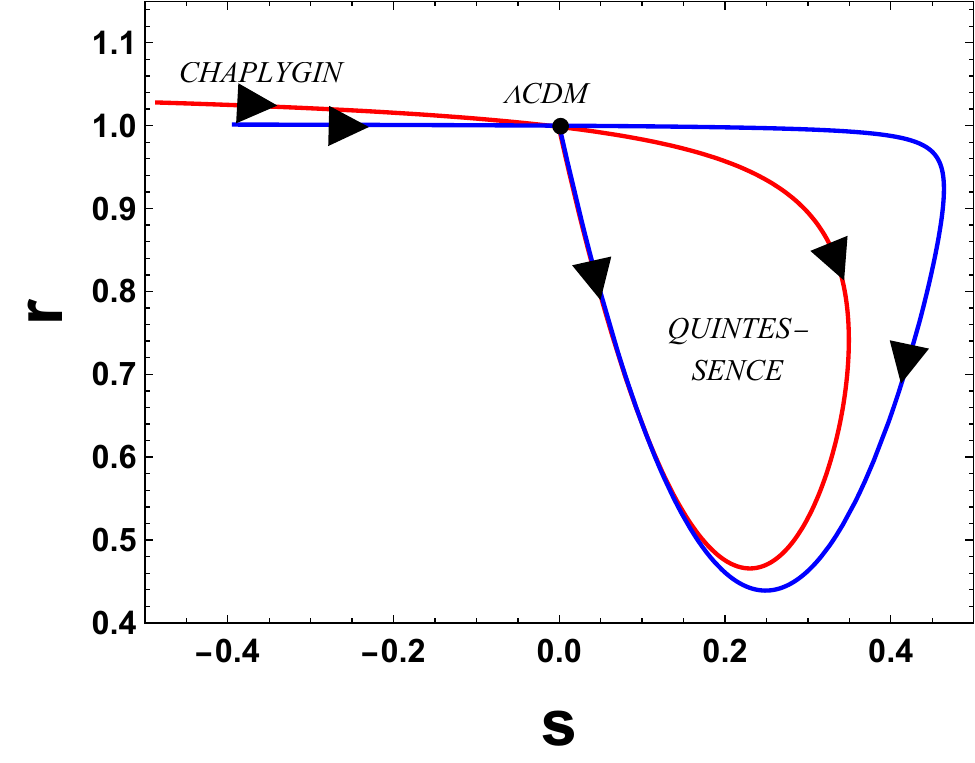}\caption{$r-s$ trajectory.} \label{Fig. viia}
\end{subfigure}
\hfill
\begin{subfigure}[b]{0.45\textwidth}
        \centering
        \includegraphics[width=\textwidth]{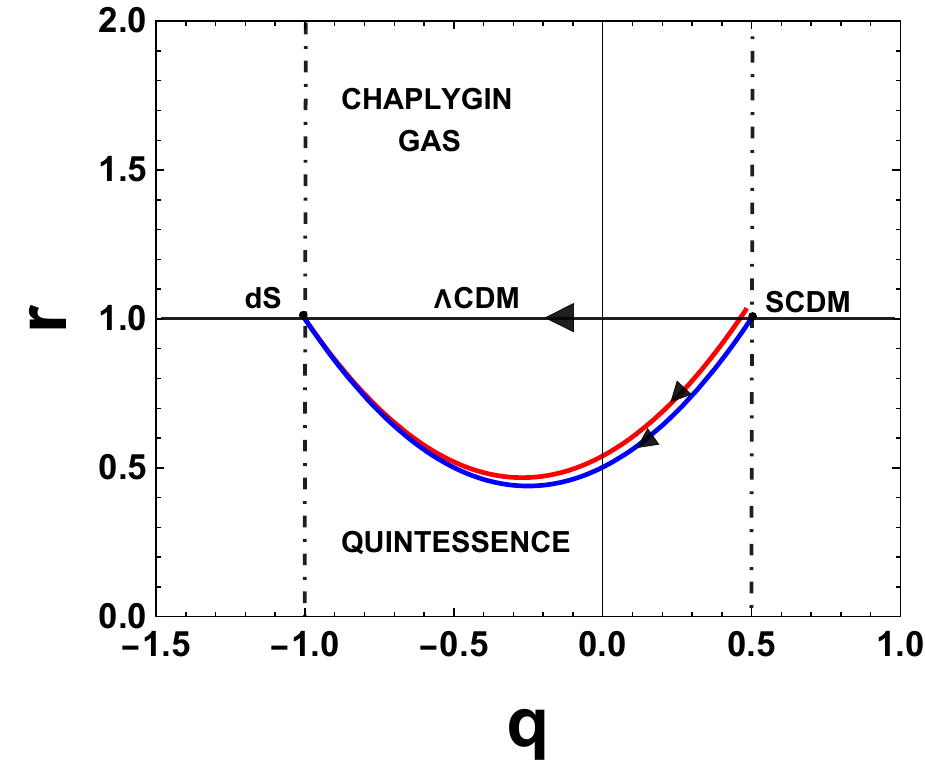}
\caption{$r-q$ trajectory.} \label{Fig. viib}
    \end{subfigure}
\caption{\justifying{Plot of the statefinder parameter for constrained values of model parameters, where the blue line represents the $OHD$ sample and the red line represents the $pantheon$ data sample.}}
    \label{Fig. vii}
\end{figure}

Specific pairs are often associated with standard dark energy models. For instance, $\{r, s\} = \{1, 0\}$ represents the $\Lambda$CDM model, whereas $\{r, s\} = \{1, 1\}$ corresponds to the standard cold dark matter (SCDM) model within the FLRW universe. The range $(-\infty, \infty)$ represents the Einstein static universe. In the $r-s$ plane, dark energy models can be classified as quintessence-like when $s>0$ and phantom-like when $s<0$. A transition from phantom to quintessence occurs when the values deviate from the standard pair $\{r, s\} = \{1, 0\}$. Fig. \eqref{Fig. viia} shows that the trajectory initially begins from the Chaplyign gas regime $(r>1, s<0)$, then enters the quintessence zone, i.e., $(r<1, s>0)$ and converges smoothly to the $\Lambda$CDM point. In Fig. \eqref{Fig. viib}, we present the temporal evolution of our model within the $\{r, q\}$ trajectory profile to extract further insights into the parameterization. In this diagnostic framework, the central solid line represents the evolution of the standard $\Lambda$CDM line, bisecting the plane into two regions: the lower half corresponds to quintessence dark energy models. In contrast, the upper half is associated with Chaplygin gas dark energy models. The trajectory initiates in the region where \(q > 0\) and \(r > 1\) (i.e., $\{q,r\}=\{0.5,1\}$) in the past, which corresponds to a matter-dominated SCDM universe. Subsequently, it transitions through the domain where \(r < 1\) and \(q < 0\) (i.e., the quintessence regime), before ultimately converging towards the de Sitter point, characterized by \(r = 1\) and \(q = -1\). Eventually, both tend to evolve like a $\Lambda$CDM universe/de-Sitter point, i.e., $\{r,s\} = \{1, 0\}$, or $\{q,r\} = \{-1, 1\}$ \cite{r108,r109,r110,kale2023transit}.

%\begin{widetext}

%\end{widetext}

\subsection{Jerk, snap, and lerk parameters}
Cosmographic methods are widely employed in DE research as they enable the study of cosmic expansion through observable parameters, bypassing the need for specific assumptions about the universe's fundamental physical structure. This model-independent approach is particularly valuable in DE studies, where theoretical assumptions about gravity and space-time, such as those in general relativity or fractal cosmology, may limit interpretation or application. Cosmographic parameters like the jerk, snap, and lerk provide insight into the rate and nature of cosmic expansion at higher orders, thus offering a detailed observational description of the universe's expansion history.
In our study, we leverage cosmography within the framework of fractal cosmology, enabling us to analyze DE behavior without committing to a singular physical model. By calculating and interpreting cosmographic parameters, we assess the fractal DE model's capacity to describe the observed accelerated expansion, aligning our analysis closely with empirical data from Observational Hubble Data (OHD) and Pantheon Supernovae samples.

The jerk $(j)$, snap $(s)$, and lerk $(l)$ parameters, which are higher-order derivatives of the deceleration parameter, provide crucial details regarding the evolution of the universe. They are shown as \cite{r111}.

%\begin{widetext}
    
\begin{equation}\label{34}
    j=(1+z) \frac{dq}{dz} + q(1+2q),
\end{equation}
\begin{equation}\label{35}
    s=-(1+z) \frac{dj}{dz}-q(2+3q),
\end{equation}
\begin{equation}\label{36}
    l=(1+z) \frac{ds}{dz}-s(3+4q).
\end{equation}

%\begin{widetext}

\begin{figure}[!h]\renewcommand{\arraystretch}{2}
\centering
\begin{subfigure}[b]{0.36\textwidth}
\centering
\includegraphics[width=\textwidth]{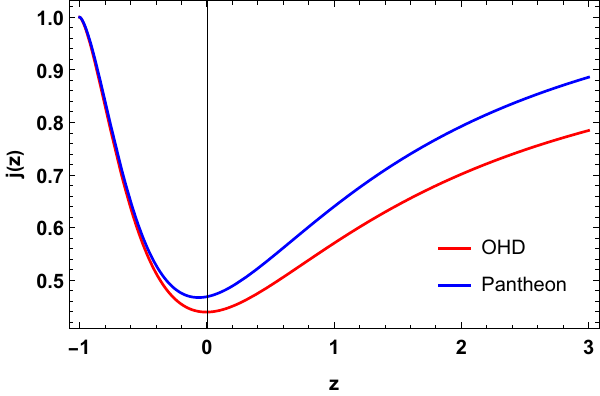}\caption{\quad Jerk parameter $(j)$ versus redshift $(z)$.} \label{viii a}
\end{subfigure}
\hfill
\begin{subfigure}[b]{0.36\textwidth}
        \centering
        \includegraphics[width=\textwidth]{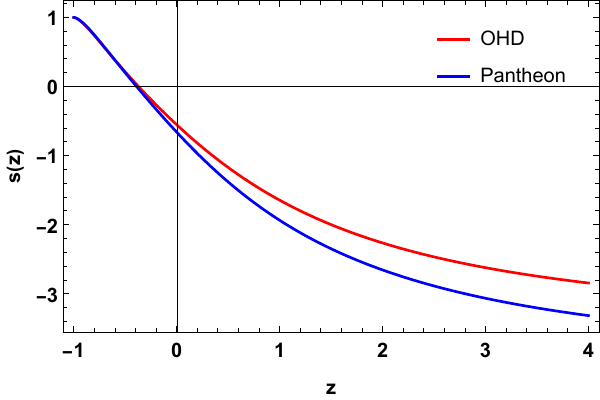}
\caption{\quad Snap parameter $(l)$ versus redshift $(z)$.} \label{viii b}
    \end{subfigure}
    \hfill
\begin{subfigure}[b]{0.36\textwidth}    \centering
\includegraphics[width=\textwidth]{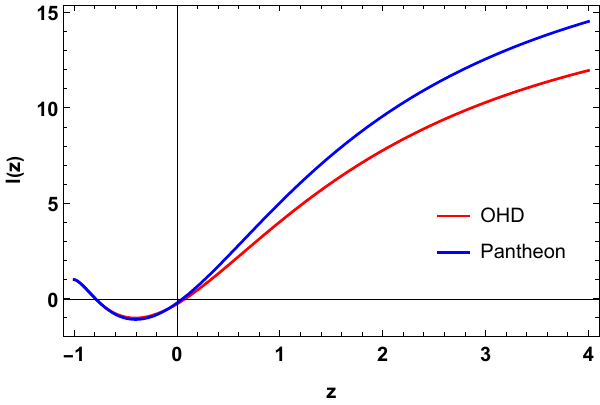}
\caption{\quad Lerk parameter $(l)$ versus redshift $(z)$.} \label{viii c}
\end{subfigure}
\caption{\justifying{Plot of the jerk, snap, and lerk parameters versus redshift $(z)$ for the $OHD$ and $pantheon$ datasets with constrained values of the model parameter $\lambda$.}}
    \label{viii}
\end{figure}
%\end{widetext}

The evolution of the deceleration parameter $(q)$ is represented by the jerk parameter $(j)$. The future is predicted via the jerk parameter since ($q$) is constrained by observation data. Furthermore, higher derivatives such as the snap $(s)$ and lerk $(l)$ parameters, teamed with the jerk parameter, offer helpful insights into the emergence of unpredictable future singularities \cite{r111}.

Figs. \eqref{viii a} and \eqref{viii c} show that the jerk and lerk parameters exhibit a decreasing trend.
These parameters remain positive throughout the evolution, indicating that the accelerated expansion of the universe is currently increasing. At $z=0$, the estimated values of $(j)$ are $j_0 = 0.439^{+0.002}_{-0.003}$ and $j_0 = 0.468^{+1.447}_{-0.086}$ for the $OHD$ and $pantheon$ samples, respectively. This lower value indicates a more gradual transition in acceleration compared with the $\Lambda$CDM model. Interestingly, $(j)$ fails to reach unity at $z=0$, which deviates from the $\Lambda$CDM model. On the other hand, in Fig. \eqref{viii b},  the snap parameter $(s)$ is observed to have increasing behavior and a negative value throughout the evolution, which supports the accelerated expansion of the universe. The $s_0$ values indicate that our model has a negative deviation from $\Lambda$CDM behavior for each data sample. The current values of $(s)$ are shown in Tab. \eqref{tab III}.

\subsection{$Om(z)$ diagnostic}

Sahni et al. \cite{r118} presented a novel $Om(z)$ diagnostic as an addition to the statefinder diagnostic. It combines the Hubble parameter $H$ and cosmic redshift $z$. This diagnosis helps to differentiate various dark energy models by tracking the slope of $Om(z)$. For a spatially flat universe, the $Om(z)$ diagnostic is defined as

\begin{equation}\label{37}
    Om(z)= \frac{E^2(z) -1}{(1+z)^3 -1},
\end{equation}
where $E(z)=H(z)/H_0$. The Hubble constant value at $z=0$ for our model is obtained as $H_0=\frac{2\mu}{1+\lambda}$; therefore, $Om(z)$ for our model is rewritten as

\begin{equation}\label{38}
    Om(z)=\frac{[1+(1+z)^{1+\lambda}]^2-4}{4z[3+z(3+z)]}.
\end{equation}

%\begin{widetext}

\begin{figure}[!h]
    \centering
    \includegraphics[width=0.45\textwidth]{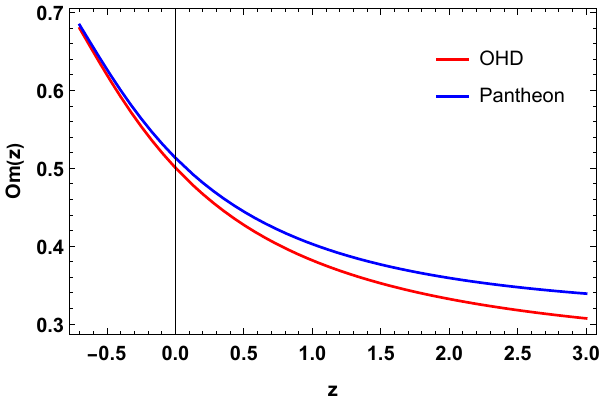}
    \caption{\justifying{Plot of $Om(z)$ versus redshift $(z)$ for the $OHD$ and $pantheon$ datasets with constrained values of the model parameter $\lambda$.}}
    \label{fig ix}
\end{figure}
%\end{widetext}
%\end{widetext}
We observe distinct O$m(z)$ values for the quintessence, phantom, and $\Lambda$CDM dark energy models. A negative slope in $Om(z)$ indicates that dark energy behaves like a quintessence (where $\omega_{eff} > -1$). Conversely, a positive slope suggests phantom-type dark energy ($\omega_{eff} <-1$), whereas a constant $Om(z)$ is characteristic of the cosmological constant, as in the $\Lambda$CDM model. 

In Fig. \eqref{fig ix}, the evolution of the $Om(z)$ diagnostic is clearly shown with a negative slope, indicating that our model is consistent with the quintessence phase of dark energy with a slowly evolving equation of state. This result aligns with the theoretical expectations of quintessence, where dark energy is dynamic and evolves, in contrast with the static nature of the cosmological constant.

\section{Energy conditions}
\label{section 6}

In modern cosmology, energy conditions are fundamental assumptions used to describe the behavior of matter and energy in the universe, particularly in the context of general relativity. These conditions provide constraints on the stress-energy tensor, which represents the distribution of matter and energy. There are several types of energy conditions, including the weak energy condition (WEC), which requires that the energy density is nonnegative for all observers, and the strong energy condition (SEC), which implies that gravity is attractive. The null energy condition (NEC) and dominant energy condition (DEC) are other important formulations. These conditions play crucial roles in identifying key theorems, such as the singularity theorems of Hawking and Penrose \cite{r113}, and are often used to test the validity of various cosmological models, including those involving dark energy, modified gravity, and the accelerating universe. Violations of certain energy conditions may point to exotic forms of energy or modifications of Einstein's theory of general relativity. These conditions are defined as \cite{r114}.
%\begin{widetext}
\begin{itemize}
   \centering
    \item Weak energy condition (WEC): $\rho_{DE} \geq 0, \rho_{DE}+P_{DE} \geq 0$.
    \item Strong energy condition (SEC): $\rho_{DE}+ P_{DE} \geq 0, \rho_{DE}+ 3P_{DE} \geq 0$.
    \item Null energy condition (NEC): $\rho_{DE}+ P_{DE} \geq 0$.
    \item Dominant energy condition: $\rho_{DE} \geq 0, \rho_{DE} \pm P_{DE} \geq 0 $.
\end{itemize}
%\end{widetext}

\begin{widetext}

\begin{figure}[!ht]
\centering
\begin{subfigure}[b]{0.31\textwidth}
\centering
\includegraphics[width=\textwidth]{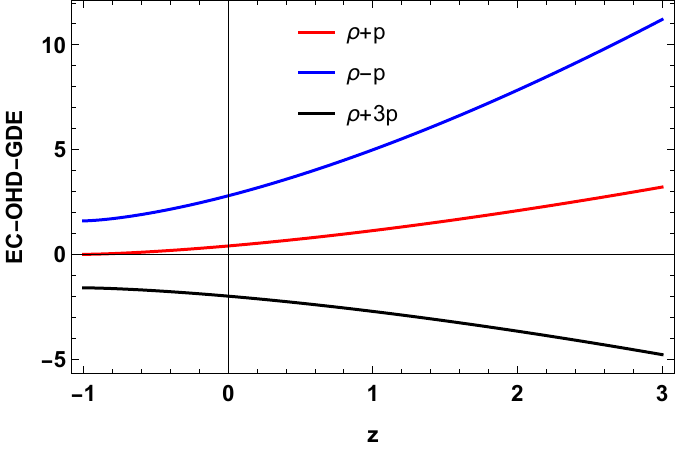}\caption{GDE model} \label{}
\end{subfigure}
\hfill
\begin{subfigure}[b]{0.33\textwidth}
        \centering
        \includegraphics[width=\textwidth]{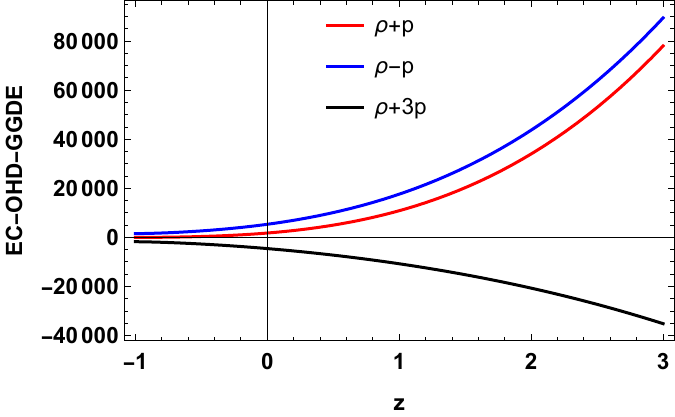}
\caption{GGDE model} \label{}
    \end{subfigure}
    \hfill
\begin{subfigure}[b]{0.34\textwidth}    \centering
\includegraphics[width=\textwidth]{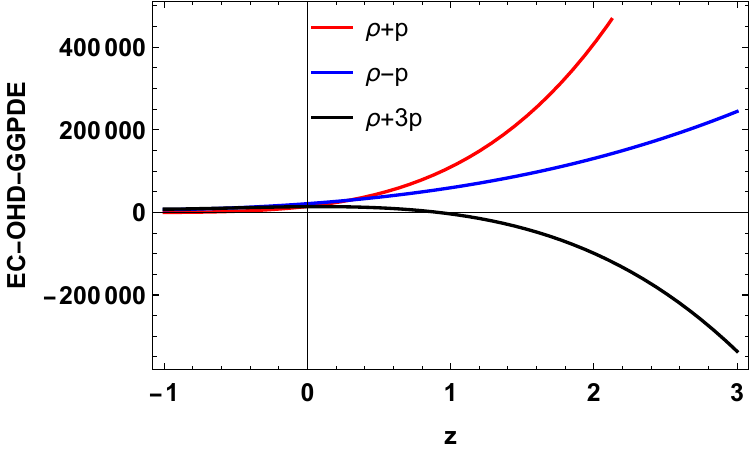}
\caption{GGPDE model} \label{}
\end{subfigure}
\caption{Plot of the energy conditions of the GDE, GGDE, and GGPDE models for the $OHD$ dataset.}
    \label{x}
\end{figure}

\begin{figure}
\centering
\begin{subfigure}[b]{0.31\textwidth}
\centering
\includegraphics[width=\textwidth]{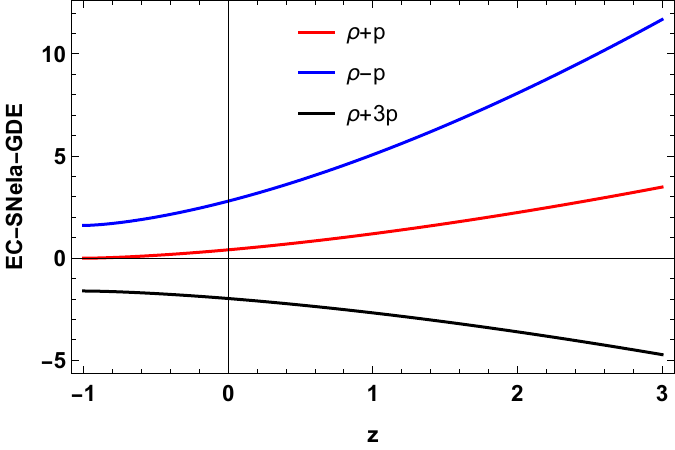}\caption{GDE model} \label{}
\end{subfigure}
\hfill
\begin{subfigure}[b]{0.33\textwidth}
        \centering
        \includegraphics[width=\textwidth]{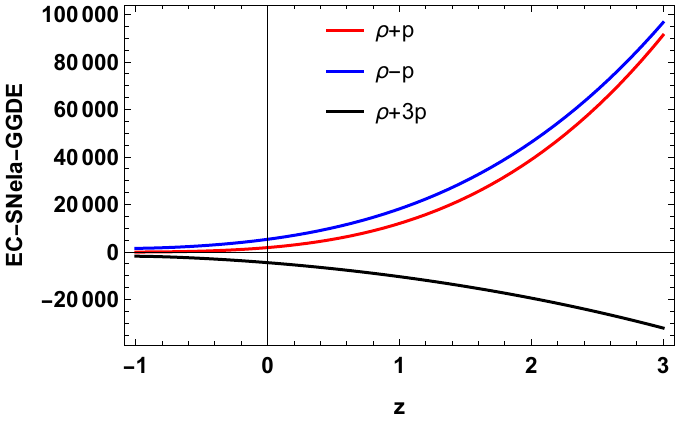}
\caption{GGDE model} \label{}
    \end{subfigure}
    \hfill
\begin{subfigure}[b]{0.34\textwidth}    \centering
\includegraphics[width=\textwidth]{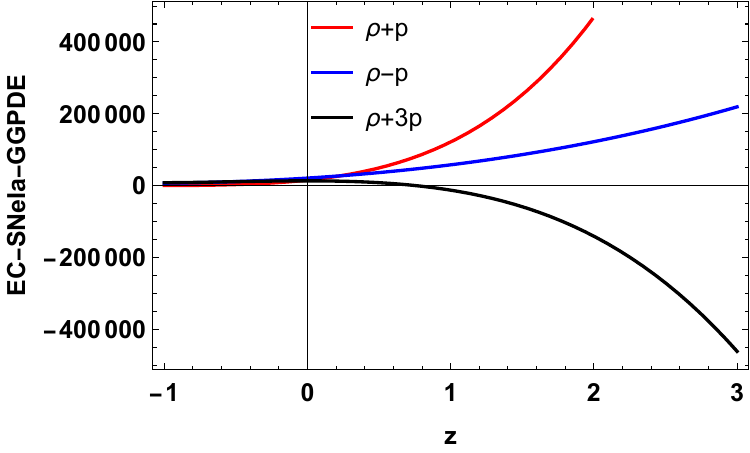}
\caption{GGPDE model} \label{}
\end{subfigure}
\caption{Plot of the energy conditions of the GDE, GGDE, and GGPDE models for the $pantheon$ dataset.}
    \label{xi}
\end{figure}

\end{widetext}
Figures \eqref{x} and \eqref{xi} depict the behavior of the energy conditions of the GDE, GGDE, and GGPDE models for constrained values of model parameters monitored by observational data samples. Each model obeys the NEC, DEC, and WEC but disobeys the SEC. Specifically, under fractal cosmology, the SEC is crucial for explaining the attractive or repulsive nature of gravity. These arguments explain the accelerated expansion of the Universe.

\begin{widetext}

\begin{table}[!h]\renewcommand{\arraystretch}{1.5}
\centering
\begin{tabular}{c c c}
\hline
Parameters  & $OHD$   & $Pantheon$\\
\hline
$H_0$ & $63.933\pm0.053$ & $64.155^{+0.055}_{-0.043}$ \\
%\hline
$\lambda$ & $0.502 \pm0.003$ & $0.540^{+0.95}_{-0.96}$ \\
%\hline
$\mu$ & $48.014 \pm 0.056$ & $49.4 \pm 3.6$ \\
\hline
\end{tabular}
\caption{ The parametric constraining result obtained via $OHD$, and $pantheon$ data.}
\label{tab I}
\end{table}

\begin{table}[!h]\renewcommand{\arraystretch}{1.3}
\begin{center}
%\adjustbox{width=\textwidth}
{
\begin{tabular}{l c  c  c}
%\noalign{\doubleline}
\hline
$\quad \quad \omega_0$& $OHD$ & $Pantheon$ &   \\
\hline
 GDE Model & $-0.697 \pm 0.0001$ & $-0.687^{+0.270}_{-0.209}$   &  \\ 
 GGDE Model  & $ -0.595 \pm 0.0001$ & $-0.593 ^{+0.063}_{-0.049}$   &  \\ 
GGPDE Model  & $-0.4164 \pm 0.0001$ & $-0.413^{+0.090}_{-0.069}$&   \\ 
\hline

\end{tabular}
}
\caption{ Summary of the best-fit values of effective EoS parameter using the $OHD$ and $pantheon$ datasets, including the confidence levels.}
\label{tab II}
\end{center}
\end{table}

\begin{table}[!ht]
\renewcommand{\arraystretch}{1.5}
\begin{center}
\label{table1}
%\adjustbox{width=\textwidth}
{
\begin{tabular}{l c c c c c c}
%\noalign{\doubleline}
\hline
Datasets& $q_0$ & $z_{t}$ &  $j_{0}$ & $s_0$ & $l_0$ &  \\
\hline
 $OHD$ & $-0.249 ^{+0.002}_{-0.001}$ &  $0.538 \pm 0.437$  & $ 0.439^{+0.002}_{-0.003}$ & $-0.552 \pm 0.008$ & $-0.274 ^{+0.003}_{-0.004}$ &  \\ 
$Pantheon$  & $ -0.230 ^{+0.475}_ {-0.480}$ &$0.4853 \pm 0.111$    & $0.468^{+1.447}_{-0.086}$ & $-0.661 \pm 0.456$ & $-0.223\pm 0.06$ &   \\ 

\hline

\end{tabular}
}
\caption{ \justifying{Summary of the best-fit values of the model parameters and statistical analysis results for the $OHD$ and $pantheon$ datasets, including the confidence levels.}}
\label{tab III}
\end{center}
\end{table}
\end{widetext}

\newpage

\section{Conclusion}
\label{section 7}

Within the framework of fractal cosmology, we investigated the dynamic features of three dark energy models (GDE, GGDE, and GGPDE). To obtain the solutions of the field equations, we adapted the Eq. \eqref{12} suggested by R.K. Tiwari et al.\cite{r85,r86} and performed parameterization on the deceleration parameter. To obtain optimistic results, we use the $R^2$ technique [Sect. \eqref{section 3}] to obtain the best-fit values of the model parameters $\lambda$ and $\mu$. $R^2-test$ determines the proportion of variance in the dependent variable that the independent variable can explain. $R^2=1$ shows the exact fit values of the model parameters $\lambda$ and $\mu$ for both datasets. For this, we utilized the $OHD$ and $Pantheon$ samples. The constrained values of the model parameters with $1-\sigma $ confidence levels are displayed in Tab. \eqref{tab I}. Table \eqref{tab III} displays the best-fit values of cosmological parameters at present intervals.

 The best-fit plots based on the observational dataset are given in Figs. \eqref{Fig. 1} and \eqref{Fig. 2}. The values of $R^2$  for the $OHD$ and $Pantheon$ data samples are $0.93$,  and $0.9935$, respectively. $Pantheon$ has 1048 observations and gives the best fit compared with $OHD$. As shown in Tab. \eqref{tab I}, the deviation of  $OHD$  is greater than that of the $Pantheon$ datasets. This $H_0$ is $63.933$ $Km/s/Mpc$ and $64.155$ $Km/s/Mpc$ for $OHD$ and $Pantheon$ respectively, which are compatible with \cite{spergel1997age}. 

 A dedicated Sect. \eqref{section 5} has been provided for the study of cosmographic parameters. 
The evolution of the deceleration parameter in Fig. \eqref{Fig.5} indicates that our cosmology involves a transition from a decelerated phase to an accelerated phase. The transition redshift corresponds to the values of the model parameters constrained by the  $OHD$ and $Pantheon$ dataset being $z_{da}=0.583\pm 0.437$ and $z_{da}=0.485 \pm 0.111$, respectively, which are compatible with \cite{mandal2020accelerating,kale2023transit}.  Moreover, the present value of the deceleration parameter for both datasets is $q_0=-0.249 ^{+0.002}_{-0.001}, -0.230^{+0.475}_ {-0.480}$\cite{solanke2023lrs,solanke2023anisotropic}.

Figure \eqref{Fig. 6} illustrates the behavior of the $\omega_{eff}$ parameter.The GDE and GGDE models remain in the quintessence era, where dark energy dominates both in the early and late eras \cite{r106,myrzakulov2023constrained}. In contrast, $0 \leq \omega_{eff} <0.3$, indicating that the GGPDE model experienced dark matter or radiation domination in the past \cite{velasquez2019reconstruction,r104}. However, as time progresses, dark energy becomes dominant and similar to the GDE and GGDE models, the GGPDE model enters the quintessence era. The current values of the EoS parameters for each dark energy model are listed in Tab. \eqref{tab II}, which shows that the GDE models closely resemble the standard $\Lambda$CDM model at present. Moreover, all the models are expected to approach the Einstein-de Sitter model in the near future \cite{r107,r108,pati2022rip,lohakare2023analyzing,gadbail2022parametrization,gadbail2022reconstruction}.

The statefinder profiles for constrained values of model parameters through the $OHD$ and $Pantheon$ data samples are displayed in Fig. \eqref{Fig. vii}. Initially, the trajectory of the $r-s$ plane lies in the region where \( r > 1 \) and \( s < 0 \), [as shown in Fig. \eqref{Fig. viia}] which links it to the domain typically associated with Chaplygin gas. As the model progresses, it passes through a transition point at \( \{r, s\} = \{1, 0\} \). This indicates a shift from the Chaplygin gas region into the quintessence region. Ultimately, the model reaches the same fixed point, \( \{r, s\} = \{1, 0\} \), which aligns with the widely accepted $\Lambda$CDM model. On the other hand, the $r-q$ profile displayed in Fig. \eqref{Fig. viib} begins at the fixed point \( \{r, q\} = \{1, 0.5\} \), corresponding to the SCDM model. It then stays within the quintessence region throughout its evolution, eventually reaching the de Sitter point at \( \{-1, 1\} \). This suggests that the model experiences a continuous phase of cosmic acceleration, driven by quintessence-like matter. These observations align well with \cite{r108,r109,r110,kale2023transit}.

The sign of the jerk parameter [Fig. \eqref{viii a}] determines shifts in the universe's dynamics, with a positive value indicating a transition period during which the universe changes its expansion rate. Initially for the early universe for $q>0$, $j$ fails to reach unity; in contrast, as time progresses, $q<0$, $j$ attains unity, which signifies that the universe is undergoing a phase of accelerated expansion that aligns with the $\Lambda$CDM paradigm. For $z=0$, the value of $j$ is always 1 for the $\Lambda$CDM model. However, in our work, we fail to achieve this value. On the other hand, the snap parameter plays a crucial role in distinguishing between a varying dark energy component and the behavior of the cosmological constant. The rate of acceleration increases as time progresses, as shown in Fig. \eqref{viii b}. The present values of these parameters are displayed in Tab. \eqref{tab III} and hold similar arguments with \cite{r112,arora2020energy}.

Moreover, Fig. \eqref{fig ix} shows the decreasing trend i.e., the negative slope of the $Om(z)$ which indicates that the dark energy may be evolving rather than remaining constant, as the standard \(\Lambda\)CDM model suggests. In the \(\Lambda\)CDM framework, dark energy is treated as a cosmological constant resulting in a relatively flat graph. However, a negative slope in \( Om(z) \) indicates that dark energy density could be increasing as the universe expands. This suggests that dark energy might have a dynamic nature, aligning more closely with models where dark energy varies, such as quintessence or other time-dependent scenarios. Such a slope implies a deviation from \(\Lambda\)CDM, pointing to alternative explanations for cosmic acceleration. This finding has broad implications, as it suggests that the rate of cosmic expansion and the behavior of large-scale structures might differ from \(\Lambda\)CDM predictions \cite{r119,r120, bouali2023data}.

In Section \eqref{section 6}, the energy conditions of GDE, GGDE, and GGPDE are examined to assess the viability of the proposed solution. All energy conditions, except for SEC, yield positive results [refer to Figs. \eqref{x} and \eqref{xi}]. The violation of the SEC strongly indicates the accelerating nature of cosmic expansion, highlighting the transition from a decelerated era to an accelerated era. Our models show the same profiles for energy conditions, as demonstrated by the $\Lambda$CDM \cite{r115,r116,r117}.

Our model primarily explores the late-time behavior of DE within the quintessence regime, where DE drives the accelerated expansion of the universe with an EoS parameter $-1< \omega <0$. However, it is crucial to consider potential catastrophic outcomes that could arise if DE evolves differently in the future. In particular, if DE were to shift into a phantom regime ($\omega<-1$), this could lead to a big rip scenario where the universe's accelerated expansion intensifies to the point that it eventually disrupts all cosmic structures from galaxies to atomic particles. Alternatively, a transition of DE to a cosmological constant-like behavior ($\omega=-1$) could result in a big freeze, where the universe continues expanding indefinitely, with galaxies moving farther apart and the universe cooling asymptotically toward absolute zero. By incorporating these potential catastrophic fates into our model’s framework, future studies can assess whether fractal cosmology offers a unique prediction for the universe’s end states and how its predictions align with or diverge from standard models such as $\Lambda$CDM. This approach would provide a more comprehensive view of DE evolution and the long-term implications of fractal cosmology.

%\begin{equation*} \rho_{GDE}=\xi H \end{equation*}
%\begin{equation*}  p_{GDE}=-\biggl[\frac{\xi \Dot{H}}{\bigl(3H-\frac{\beta}{t}\bigr)} +\xi H \biggr] \end{equation*}

%\begin{equation*} \omega_{GDE} =-\biggl[\frac{\Dot{H}}{H \bigl(3H-\frac{\beta}{t}\bigr)} +1 \biggr] \end{equation*}  

%\begin{equation} \rho_{GDE} + p_ {GDE} = - \frac{\xi \Dot{H}}{\bigl(3H-\frac{\beta}{t}\bigr)} \end{equation}

%\begin{equation} \rho_{GGDE}=\xi H +\eta H^2 \end{equation}

%\begin{equation} p_{GGDE}=- \biggl[\frac{\xi \Dot{H} +2\eta H \Dot{H}}{3H-\frac{\beta}{t}}+\xi H + \eta H^2 \biggr] \end{equation}

%\begin{equation} \omega_{GGDE} = - \biggl[ \frac{\xi \Dot{H} +2\eta H \Dot{H}}{\bigl(3H-\frac{\beta}{t}\bigr)\bigl(\xi H + \eta H^2 \bigr)} +1 \biggr] \end{equation}

%\begin{equation}  \rho_{GGDE} + p_{GGDE} = - \biggl[\frac{\xi \Dot{H} +2\eta H \Dot{H}}{3H-\frac{\beta}{t}}\biggr] \end{equation}

%\begin{equation}  \rho_{GGPDE}= \bigl(\xi H +\eta H^2 \bigr)^\psi \end{equation}

%\begin{equation} p_{GGPDE}=- \biggl[\frac{\psi\big(\xi H + \eta H^2\big)^\psi \big(\xi \Dot{H} +2\eta H \Dot{H}\big)}{\big(3H-\frac{\beta}{t}\big)\big( \xi H + \eta H^2 \big)}+\big(\xi H + \eta H^2\big)^\psi \biggr] \end{equation}

%\begin{equation} \omega_{GGPDE} = - \biggl[ \frac{\psi\big(\xi \Dot{H} +2\eta H \Dot{H}\big)}{\bigl(3H-\frac{\beta}{t}\bigr)\bigl(\xi H + \eta H^2 \bigr)} +1 \biggr] \end{equation}

%\begin{equation} \rho_{GGPDE}+ p_{GGPDE}=- \biggl[\frac{\psi\big(\xi H + \eta H^2\big)^\psi \big (\xi \Dot{H} +2\eta H \Dot{H}\big)}{\big(3H-\frac{\beta}{t}\big)\big( \xi H + \eta H^2 \big)} \biggr] \end{equation}

\bibliographystyle{rsc}
\bibliography{clean_main_R2}
\end{document}